\documentclass[aps,eprint,superscriptaddress,twocolumn,nofootinbib]{revtex4-2}
\usepackage{amsmath}
\usepackage{amsthm}
\usepackage{mathtools}
\usepackage{hyperref}
\usepackage{amssymb}
\usepackage{physics}
\usepackage{graphicx}
\graphicspath{ {images/} }
\usepackage{float}

\usepackage{xcolor}

\usepackage{comment}

\usepackage{setspace} 

\usepackage[section]{placeins} 

\usepackage{pdfpages}
\makeatletter
\AtBeginDocument{\let\LS@rot\@undefined}
\makeatother
\usepackage[portrait, margin=0.5in]{geometry}

\newcommand{\p}[2]{\ensuremath{\frac{\partial #1}{\partial #2}}} 
\newcommand{\const}{\mathrm{const.}} 

\newcommand{\beq}{\begin{equation}}
\newcommand{\eeq}{\end{equation}}

\begin{document}
	
\title{Defect absorption and emission for $p$-atic liquid crystals on cones}
\author{Farzan Vafa}
\affiliation{Center of Mathematical Sciences and Applications, Harvard University, Cambridge, MA 02138, USA}
\author{Grace H. Zhang}
\affiliation{Department of Physics, Harvard University, Cambridge, MA 02138, USA}
\author{David R. Nelson}
\affiliation{Department of Physics, Harvard University, Cambridge, MA 02138, USA}
\date{\today}
	
\begin{abstract}
		
We investigate the ground state configurations of $p$-atic liquid crystals on fixed curved surfaces. We focus on the intrinsic geometry and show that isothermal coordinates are particularly convenient as they explicitly encode a geometric contribution to the elastic potential. In the special case of a cone with half-angle $\beta$, the apex develops an effective topological charge of $-\chi$, where $2\pi\chi = 2\pi(1-\sin\beta)$ is the deficit angle of the cone, and a topological defect of charge $\sigma$ behaves as if it had an effective topological charge $Q_\mathrm{eff} = (\sigma - \sigma^2/2)$ when interacting with the apex. The effective charge of the apex leads to defect absorption and emission at the cone apex as the deficit angle of the cone is varied.
		
For total topological defect charge 1, e.g. imposed by tangential boundary conditions at the edge, we find that for a disk the ground state configuration consists of $p$ defects each of charge $+1/p$ lying equally spaced on a concentric ring of radius $d = (\frac{p-1}{3p-1})^{\frac{1}{2p}} R$, where $R$ is the radius of the disk. In the case of a cone with tangential boundary conditions at the base, we find three types of ground state configurations as a function of cone angle: (1) for sharp cones, all of the $+1/p$ defects are absorbed by the apex; (2) at intermediate cone angles, some of the $+1/p$ defects are absorbed by the apex and the rest lie equally spaced along a concentric ring on the flank; and (3) for nearly flat cones, all of the $+1/p$ defects lie equally spaced along a concentric ring on the flank. Here the defect positions and the absorption transitions depend intricately on $p$ and the deficit angle which we analytically compute. We check these results with numerical simulations for a set of commensurate cone angles and find excellent agreement.
		
\end{abstract}
	
\maketitle
	
	
\section{Introduction}
	
Two-dimensional liquid crystals with $p$-fold rotational symmetry, denoted ``p-atics'', are ubiquitous in nature. One well-studied example is the hexatic ($p=6$) phase, an intermediate phase that can appear when isotropic two-dimensional liquids freeze into $2d$ crystals~\cite{halperin1978theory,nelson1979dislocation} within the KTHNY defect-mediated melting scenario~\cite{kosterlitz1972long,kosterlitz1973ordering,young1979melting}. Hexatics may be of some biological importance, because they have appeared in recent computational models of epithelial monolayers~\cite{li2018role} and because they arise as an intermediate phase of lipid bilayers (see Ref.~\cite{korolev2008defect} and references therein). Continuous hexatic-to-crystal transitions, as found in experiments for lipid vesicles in Ref.~\cite{dimova2000pretransitional}, may be especially important, as they are accompanied by a tunable, continuously diverging $2d$ shear viscosity~\cite{nelson1979dislocation}.
Another well-studied example is thermotropic liquid crystals, where frequently a nematic ($p=2$) phase appears~\cite{gennes1993the}. Liquid crystalline $p$-atics have also been realized in colloidal systems, including monolayers of sedimented colloidal hard spheres in the hexatic phase~\cite{thorneywork2017two}, triatic ($p=3$) colloidal platelets~\cite{zhao2012local}, and possibly tetratic ($p=4$) suspensions of colloidal cubes~\cite{loffler2018phase}. Although one might expect that steric repulsions could produce local antiferromagnetic order for hard triangles and pentagons~\cite{mietke2022anyonic}, longer range interactions could induce these objects to align ferromagnetically, which is what we assume for $p=3$ and $p=5$ in this paper.
	
Order characteristic of $p$-atics has also been studied in the context of active matter~\cite{marchetti2013hydrodynamics}. Examples of active polar fluids ($p=1$), also known as Toner-Tu fluids~\cite{vicsek1995novel,toner1995long,toner1998flocks,toner2005hydrodynamics}, include bacterial suspensions~\cite{wensink2012meso}, groups of animals such as bird flocks~\cite{toner1995long}, and Quincke rotors~\cite{bricard2013emergence}; examples of active nematics ($p=2$)~\cite{simha2002hydrodynamic,doostmohammadi2018active} include cell sheets~\cite{duclos2017topological,kawaguchi2017topological,saw2017topological,blanch2018turbulent}, suspensions of cytoskeletal filaments and associated motor proteins~\cite{sanchez2012spontaneous,keber2014topology,kumar2018tunable}, bacteria collectives~\cite{doostmohammadi2016defect,nishiguchi2017long-range,copenhagen2020topological}, vibrated granular rods~\cite{narayan2007long}, and developing organisms~\cite{maroudas2020topological}; finally, the tissue of the brine shrimp \emph{Parhyale hawaiensis} during development provides an example of a tetratic ($p=4$) order~\cite{cislo2021active}.
	
An elastic description of $p$-atics was employed for $p=6$ hexatics on fluctuating surfaces in~\cite{nelson1987fluctuations} and later refined in~\cite{lubensky1992orientational,park1996topological}. Much work on curved surfaces has focused on effects of extrinsic geometry, such as how the surface is embedded in three dimensions, and effects due to the mean curvature~\cite{bowick2009two}. Here, we find it convenient to focus on simpler, but still quite challenging effects of intrinsic geometry, and use isothermal coordinates, as recently done in the context of morphogenesis of an active nematic~\cite{vafa2021active}. We are interested in the ground state configurations of liquid crystals on curved surfaces, in particular a cone, given constraints on the total topological charge of the defects. Ground state defect configurations for the cases of flat plane, hollow cylinder, sphere, and torus were derived in Ref.~\cite{lubensky1992orientational}. These geometries, however, are smooth and lack curvature singularities such as sharp points or ridges, characteristic of imperfect surfaces. In contrast, we study cones, the simplest example of a curvature singularity. The interaction between $p$-atic order and curved substrates has been studied in~\cite{park1996topological,vitelli2004anomalous} and it has been shown that curvature gives rise to an effective topological charge density.\footnote{Related effects have been noted in quantum Hall states for electrons on cones~\cite{wen1992shift,biswas2016fractional}.} In the special case of a cone, this would correspond to negative topological charge concentrated at the apex, and a simple argument was recently presented in~\cite{zhang2022fractional} for the case of free boundary conditions. We re-derive this induced charge result of Vitelli and Turner~\cite{vitelli2004anomalous} and use it to determine the ground state defect configuration with a fixed number of $+1/p$ defects, which appear naturally when tangential boundary conditions are imposed at the cone base. In the ground state, we find that the cone apex absorbs defects until the net topological charge at the apex becomes positive, and the remaining defects lie equally spaced on a ring optimally positioned, as a function of the cone angle, between the apex and the boundary. We derive both these transitions and the flank defect positions, which depend intricately on the deficit angle and the charges of the defects, and find excellent agreement with numerical simulations for a set of commensurate cone angles.
	
This paper is organized as follows. We begin in Sec.~\ref{sec:isothermal} with a review of isothermal coordinates, essential for our formalism. Although we focus on cones, spheres and tori are mentioned briefly to provide context. In Sec.~\ref{sec:model}, we review the formalism of $p$-atics on curved surfaces using isothermal coordinates and set up the computation of the free energy. By evaluating the free energy in Sec~\ref{sec:F}, in analogy to electrostatics, we show that topological defects interact with each other via a two-dimensional Coulombic interaction, and that there is a geometric contribution to the potential. In particular, the cone apex develops an effective negative topological charge of $-\chi$ where $\chi = 1-\sin \beta$, with $\beta$ being the half cone angle (see Fig.~\ref{fig:coordinates}c), and $2\pi\chi$ is the deficit angle of the cone. Moreover, a topological defect of charge $\sigma$, when interacting with the apex, develops an effective charge $Q_\mathrm{eff} = \sigma-\sigma^2/2$, as originally found in Ref.~\cite{vitelli2004anomalous}. In Sec.~\ref{sec:groundStates}, we describe defect absorption and emission at the cone apex, with transitions and flank defect positions depending intricately on the deficit angle of the cone and the defect charges, and find excellent agreement between these analytical results and numerical energy minimizations of lattice models laid down on cones with special commensurate curvatures that allow precise computations~\cite{zhang2022fractional}. We conclude in Sec.~\ref{sec:conclusion} by reviewing our results and suggesting future directions of research, including dynamics of active topological defects on curved surfaces, alternative boundary conditions, and analogous phenomena involving grain boundaries on cones. Some of the technical details are relegated to Appendices~\ref{app:BC}-\ref{app:ground}. 
	
\section{Isothermal coordinates primer}
\label{sec:isothermal}
	
Since our formulation is based on isothermal coordinates, we introduce them from the outset. From work dating back to Gauss~\cite{gauss1822on}, we know that in two dimensions it is always possible to choose local complex coordinates $z$ and $\bar z$, known as isothermal (or conformal) coordinates, such that the metric can be written as,
\beq 
    ds^2 = g_{z\bar z} dz d\bar z + g_{\bar z z} d\bar z dz = 2g_{z\bar z}|dz|^2 = e^{\varphi(z,\bar z)}|dz|^2 .
\eeq
Note that in these coordinates, $g^{z\bar z}$ and $g^{\bar z z}$ can be read off from the off-diagonal components of the inverse metric
\beq g^{-1} = 
    \begin{pmatrix}
        0 & 2e^{-\varphi(z, \bar z)} \\
        2e^{-\varphi(z, \bar z)} & 0
    \end{pmatrix} .
\eeq
Upon writing $z = x+iy$, $\bar z = x - iy$, we also have
\beq 
    ds^2 = e^{\varphi(x,y)}(dx^2 + dy^2) .
\eeq
Thus, the metric is conformally flat, i.e. proportional to the identity matrix, where $e^{\varphi}$, known as the conformal factor, represents a position-dependent isotropic stretching. We show in Sec.~\ref{sec:F} that we can interpret $-\varphi$ as a geometric contribution to the defect potential, and thus call $\varphi$ the geometric potential. For a more detailed presentation of isothermal coordinates, see, for example, Refs~\cite{david2004geometry} and~\cite{turner2010vortices}.
	
In complex conformal coordinates, the only nonzero Christoffel symbols are
\beq 
    \Gamma^z_{zz} = \partial \varphi, \qquad \Gamma^{\bar z}_{\bar z \bar z} = \bar\partial \varphi, 
\eeq
where the holomorphic partial derivatives are denoted as $\partial \equiv \partial_z = \frac{1}{2}\left(\p{}{x} - i \p{}{y}\right)$ and $\bar\partial \equiv \partial_{\bar z} = \frac{1}{2}\left(\p{}{x} + i \p{}{y}\right)$. 
The Laplacian $\nabla^2$ acting on a scalar $f$ is given by
\beq
    \nabla^2 f \equiv g^{z\bar z}\partial\bar\partial f +  g^{\bar z z}\bar\partial\partial f = 2 g^{z\bar z}\partial\bar\partial f = 4e^{-\varphi} \partial\bar\partial f
\eeq
As an aside, we note that in analogy to the heat equation, coordinates $z$ and $\bar z$ are harmonic, i.e., they satisfy $\nabla^2 z = \nabla^2 \bar z = 0$, and so constant coordinate lines are ``isotherms'', hence the name ``isothermal''.
Note also that the Gaussian curvature $K$ is given in terms of $\varphi$ by
\beq 
    K = -\frac{1}{2}\nabla^2\varphi =-2e^{-\varphi}\partial\bar\partial\varphi . \label{eq:K}
\eeq
Note finally the property of holomorphic derivatives that $\partial_z f(\bar z) = \partial_{\bar z} f(z) = 0$.
We now provide a few examples.
	
\subsection{Cone}
	
\begin{figure}
	\centering
	\includegraphics[width = 1\columnwidth]{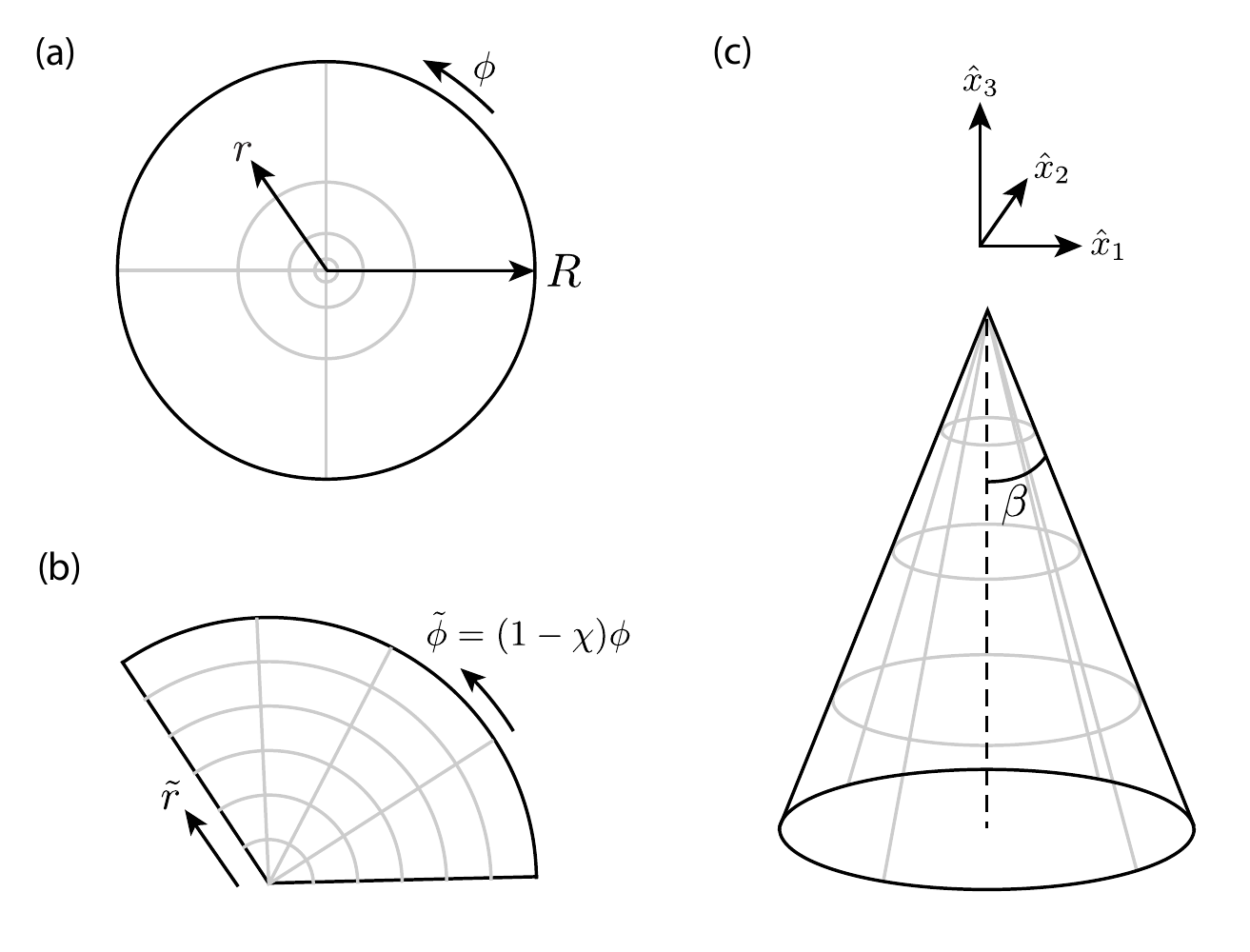}
	\caption{Schematic of the three coordinate systems for cone used in this paper: (a) Our preferred isothermal coordinates $z = re^{i\phi}$, which can be viewed as the result of squashing a cone into a plane, in a way that preserves the azimuthal angle $\phi$, $0 \le \phi < 2\pi$. Here $R$ is the maximum radius down the cone flanks in our isothermal coordinate system. (b) The also useful $\tilde z$ coordinates, the result of isometrically cutting open and unrolling a cone into a plane, so that the resulting azimuthal angle is $\tilde\phi$, $0 \le \phi < 2\pi(1-\chi)$. (c) Three-dimensional Cartesian coordinates $x_i$, where $\beta$ is the cone half-angle.}
	\label{fig:coordinates}
\end{figure}
	
On the surface of a cone, the geometric potential $\varphi$ and the metric are given by,
\beq 
    \varphi = -\chi \ln (z \bar z), \qquad ds^2 = |z|^{-2\chi} |dz|^2, \label{eq:coneMetric}
\eeq
where we will show that $2\pi\chi$ is the deficit angle. To do so, we first go to a new coordinate system $\tilde z = \tilde r e^{i\tilde\phi}$ (see Fig.~\ref{fig:coordinates}b), where the metric can be made flat with no stretching via the change of coordinates
\beq 
    \tilde z = \frac{z^{1-\chi}}{1 - \chi} ,\label{eq:ztilde}
\eeq
which leads to
\beq 
    ds^2 = |d\tilde z|^2 .
\eeq
This is a flat metric except at the origin, where $\tilde z$ is not well-defined. To understand the geometry near the origin, note that Eq.~\eqref{eq:ztilde} gives the angle $\tilde \phi$ corresponding to $\tilde z = \tilde r e^{i \tilde \phi}$, in terms of the original complex conformal coordinate $z = re^{i\phi}$, as
\beq 
    \tilde\phi = (1-\chi)\phi .
\eeq
Thus, the range of polar angle $\tilde \phi$ in $\tilde z$ coordinates is $0\le\tilde\phi <2\pi(1-\chi)$, so this geometry has a conical singularity with deficit angle $2\pi\chi$. 
	
We now show that in terms of the cone half-angle $\beta$, $\chi = 1-\sin\beta$. To show this relation, we go to one final coordinate system $x_i$ (see Fig.~\ref{fig:coordinates}c), which embeds the cone in three dimensions, $\vec x = \vec x(r,\phi)$, with
\begin{align}
	x_1 &= r^{1-\chi}\cos\phi\\
	x_2 &= r^{1-\chi}\sin\phi \\
	x_3 &= -\frac{\sqrt{1 - (1 - \chi)^2}}{1-\chi}r^{1-\chi} .
\end{align}
With this change of coordinates, the metric can be expressed as
\beq 
    ds^2 = |z|^{-2\chi}|dz|^2 = dx_1^2 + dx_2^2 + dx_3^2.
\eeq
Notice that since
\beq 
    x_3 = -\frac{\sqrt{1 - (1 - \chi)^2}}{1-\chi} \sqrt{x_1^2 + x_2^2} = -\cot\beta  \sqrt{x_1^2 + x_2^2}, 
\eeq
where $\beta$ is the cone half angle, it follows that
\beq 
    \chi = 1-\sin\beta, \label{eq:chi}
\eeq
thus relating the deficit angle $2\pi\chi$ to the cone half angle $\beta$.

\subsection{Sphere}
	
For the surface of a unit sphere, $\varphi$ is
\beq 
    \varphi = 2\ln \frac{2}{1 + |z|^2}, 
\eeq
which is equivalent to the stereographic projection, and can be viewed as the mapping of the complex plane $z$ onto the points $(x_1, x_2, x_3)$ on the surface of the unit sphere in $\mathbb R^3$, via
\begin{align}
	x_1 &= \frac{z+\bar z}{1 + |z|^2}\\
	x_2 &= \frac{1}{i}\frac{z-\bar z}{1 + |z|^2}\\
	x_3 &= \frac{|z|^2-1}{1 + |z|^2}.
\end{align}
Hence the metric is
\beq 
    ds^2 = dx_1^2 + dx_2^2 + dx_3^2 = \frac{4}{(1 + |z|^2)^2}|dz|^2 \equiv e^{\varphi(z,\bar z)}|dz|^2
\eeq
and, using Eq.~\eqref{eq:K}, the Gaussian curvature is computed to be $K=1$.
	
\subsection{Torus}
As the last example, we consider the standard torus $T^2$ in $\mathbb R^3$, parametrized by:
\begin{align}
	x_1 &= (R_1 + R_2\cos\theta_2)\cos\theta_1 \\
	x_2 &= (R_1 + R_2\cos\theta_2)\sin\theta_1 \\
	x_3 &= R_2\sin\theta_2,
\end{align}
where $\theta_i$ ($0 \le \theta_i < 2\pi$) is the periodic angular variable of circle of radius $R_i$, with $R_1 > R_2$. Following Ref.~\cite{bowick2004curvature}, we now express $T^2$ in isothermal coordinates. Let $r = R_1/R_2$. Then, on making the following complex change of coordinates
\beq 
    z = \frac{1}{2\pi}\left(\phi_1 + \frac{i}{\sqrt{r^2 - 1}}\phi_2\right),
\eeq
where $\phi_1$ and $\phi_2$ are given by
\begin{align}
	\phi_1 &= \theta_1\\
	\frac{r\cos\phi_2 - 1}{r - \cos\phi_2} &= \cos \theta_2 ,
\end{align}
the metric becomes
\beq 
    ds^2 = e^{\varphi} |dz|^2,
\eeq
where
\begin{align}
	\varphi &= 2\ln(2\pi R_2) + 2\ln\left(\frac{r^2-1}{r - \cos\phi_2}\right) \nonumber \\
	&=2\ln(2\pi R_2) + 2\ln\left(\frac{r^2-1}{r - \cos\left[\frac{1}{i}\pi \sqrt{r^2 - 1}(z - \bar z)\right]}\right) .
\end{align}
In terms of $\tau = \frac{i}{\sqrt{r^2 - 1}}$, the isothermal coordinate $z$ is identified with its shifts by $1$ and $\tau$ (forming a parallelogram on the complex plane), i.e., $z\sim z+1 \sim z+\tau$. Here $\tau$ is known as the parameter specifying the ``complex structure'' of the torus~\cite{nakahara2003geometry}.  Using Eq.~\eqref{eq:K}, the Gaussian curvature is computed to be (with, again, $r = R_1/R_2$)
\beq 
    K = \frac{1}{R_2^2}\frac{r\cos\phi_2 - 1}{r^2 - 1} = \frac{1}{R_2^2}\frac{r\cos\left[\frac{1}{i}\pi \sqrt{r^2 - 1}(z - \bar z)\right] - 1}{r^2 - 1}.
\eeq
	
\section{Minimal model}
\label{sec:model}
	
For recent discussions of $p$-atic tensor order parameters in $2d$ flat space, see Refs.~\cite{giomi2021hydrodynamic,giomi2021longranged}. Theories of $p$-atics on \emph{curved} surfaces were previously formulated in Ref.~\cite{park1996topological}. For this work to be self-contained, we review the formalism and recast it in terms of isothermal coordinates, which will prove to be a powerful method. Following the presentation in Ref.~\cite{vafa2021active} and as described in Sec.~\ref{sec:isothermal}, we work with complex isothermal coordinates $z$ and $\bar z$. Let $\mathbf{Q}$ be the $p$-atic tensor, a traceless real symmetrized rank-$p$ tensor. Since $\mathbf{Q}$ is traceless (contraction of any pair of indices vanishes), $\mathbf{Q}$ has only two non-zero components $Q \equiv Q^{z\ldots z}$ and $\bar Q \equiv Q^{\bar z \ldots \bar z}$, where here ellipses denote $p$ copies. Also, by reality, $Q = (\bar Q)^*$. For ease of notation, let $\nabla \equiv \nabla_z$ denote the covariant derivative with respect to $z$ and $\bar\nabla \equiv \nabla_{\bar z}$ denote the covariant derivative with respect to $\bar z$. Explicitly, covariant derivatives of the $p$-atic tensor are
\beq 
    \nabla Q = \partial Q + p (\partial \varphi) Q, \qquad \bar \nabla Q = \bar \partial Q. \label{eq:nabla}
\eeq
Results for a cone with half-angle $\beta$ follow from substituting $\varphi = -(1-\sin\beta)\ln(z \bar z)$ in Eq.~\eqref{eq:nabla}.
	
To provide an intuitive explanation for the asymmetric form of the two covariant derivatives written above in Eq.~\eqref{eq:nabla}, note that Eq.~\eqref{eq:nabla} looks like $Q$ carries charge $p$ under a $U(1)$ vector potential, 
\beq 
    A_z = i\partial\varphi .
\eeq
Indeed, the rotation group in two dimensions is $SO(2)$, and is gauged by a geometric field corresponding to curved geometry (known as the spin connection)~\cite{david2004geometry}, which in holomorphic coordinates splits into two $U(1)$ gauge fields,
\beq 
    (A_z, \bar A_{\bar z}) = (i\partial\varphi, -i\bar\partial\varphi)
\eeq
The charge of the fields depend on the number of $z$ and $\bar z$ indices. In particular, $Q^{z\ldots z}$ carries charge $(p,0)$ (because it has $p$ $z$-indices and no $\bar z$-indices), and similarly $Q^{\bar z \ldots \bar z}$ carries charge $(0,p)$. This explains that in Eq.~\eqref{eq:nabla}, since $Q^{z\ldots z}$ does not carry any $\bar z$ charge, it does not couple to $\bar A_{\bar z}$, thus explaining the asymmetry in the above formulae (Eq.~\eqref{eq:nabla}). Note that the field strength of this $U(1)$ gauge field, defined as
\beq 
    F_{z\bar z} \equiv \frac{1}{2i}(\partial \bar A_{\bar z} - \bar\partial A_z) = - \partial\bar\partial\varphi = R_{z\bar z} ,
\eeq
is nothing but the Ricci curvature~\cite{nakahara2003geometry}.

To keep the model simple, in a way that corresponds to the one Frank constant approximation in nematic liquid crystals~\cite{gennes1993the}, and to the Maier-Saupe lattice model used in our numerical calculations, we decouple the rotation symmetry of the $p$-atic degrees of freedom from the local rotational invariance in space. The only elastic terms are then $g_{z \bar z}^{p-1}\nabla Q \bar\nabla \bar Q$ and $g_{z \bar z}^{p-1}\bar\nabla Q \nabla \bar Q$, where we recall that $g_{z\bar z} = g_{\bar z z} = \frac{1}{2}e^{\varphi(z,\bar z)}$ and $g_{zz} = g_{\bar z \bar z} = 0$. Then, our simplified free energy can be written as
\beq 
    \mathcal F = 2^{p-1} \int d^2z \sqrt{g}[K |\nabla Q|^2 + K' |\bar \nabla Q|^2 + \epsilon^{-2} (1 - c|Q|^2)^2],\label{eq:minimal}
\eeq
where
\begin{align}
	|\nabla Q|^2 &= g_{z \bar z}^{p-1}\nabla Q \bar\nabla \bar Q \nonumber\\ 
	|\bar \nabla Q|^2 &= g_{z \bar z}^{p-1}\bar\nabla Q \nabla \bar Q \nonumber\\ 
	|Q|^2 &= g_{z \bar z}^p Q \bar Q.
\end{align}
Here $K,K' >0$ are elastic terms in the spirit of the one-Frank-constant approximation (the $K$ and $K'$ terms are equivalent in flat space), and the last term governs the amplitude of the $p$-atic order parameter, with $\epsilon$ controlling the microscopic $p$-atic correlation length. We take $c=2^p$, a normalization we choose without loss of generality.
	
We now determine $Q$ by minimizing the free energy. Deep in the ordered limit ($\epsilon \ll 1$), we have
\beq 
    2^p|Q|^2 = 1 .
\eeq
The substitution $Q^{z\ldots z} = S^{z\ldots z}e^{i\gamma} = Se^{i\gamma}$, which endows our tensor order parameter with a phase $\gamma = p\theta$, where $\theta$ is the angle the $p$-atic molecule makes with a local reference axis, then leads to
\beq 
    S = (2g_{z\bar z})^{-p/2} = e^{-\frac{p}{2}\varphi}.\label{eq:S}
\eeq
Upon inserting $\varphi = -\chi \ln (z \bar z)$ into Eq.~\eqref{eq:S}, we see that the $p$-atic order parameter amplitude $S$, in isothermal coordinates, vanishes like a power law near the cone apex, $S \sim |z|^{p \chi}$, as if near a defect core in flat space. However, the contribution of the polynomial part of the free energy vanishes away from the core apex and any defect cores, so the free energy simplifies to\footnote{Provided we introduce a phenomenological defect core energy $E_\mathrm{c}$, we could have instead started with Eq.~\eqref{eq:FSigmaModel} in combination with the constraint Eq.~\eqref{eq:S} .}
\beq 
    \mathcal F = 2^{p-1} \int d^2z \sqrt{g}[K|\nabla Q|^2 + K'|\bar\nabla Q|^2] . \label{eq:FSigmaModel}
\eeq
	
By integration by parts, it is easy to show that the $K$ and $K'$ terms differ only by a term proportional to $R|Q|^2$~\cite{vafa2021active}, where $R$ is the scalar curvature. Near the minimum of the potential, where $|Q|^2=1$, this difference becomes a Gauss-Bonnet term which is a a total derivative and thus topological. Thus, the $K$ and $K'$ terms are equivalent deep in the ordered limit that we consider in this paper. 
	
Upon substitution of $\mathbf{Q}$ (with the amplitude $S$ given by Eq.~\eqref{eq:S}) into Eq.~\eqref{eq:FSigmaModel}, the free energy $\mathcal F$ reduces to
\begin{align}
    \mathcal F &=  (K+K')\int d^2z\left(\left(\frac{p}{2}\right)\partial\varphi + i\partial\gamma\right)\left(\left(\frac{p}{2}\right)\bar\partial\varphi - i\bar\partial\gamma\right) \nonumber \\
    & = J\int d^2z\left|\partial\gamma-i\left(\frac{p}{2}\right)\partial\varphi\right|^2 ,  \label{eq:F}
\end{align}
where we have used
\begin{align}
	\nabla Q^{z\ldots z} &= \left(\frac{p}{2}\partial\varphi + i\partial\gamma\right)Q\\
	\bar\nabla Q^{\bar z\ldots \bar z} &= \left(-\frac{p}{2}\bar\partial\varphi + i\bar\partial\gamma\right)Q
\end{align}
and where $J = K + K'$. 
Note that Eq.~\eqref{eq:F} is much simpler than Eq.~\eqref{eq:FSigmaModel}. \emph{The only dependence of the free energy on the geometry is through $\partial\varphi$: there are no factors such as $e^{\varphi}$ or $\sqrt{g}$, which in two dimensions cancel due to conformal symmetry.} The transparency and simplicity of Eq.~\eqref{eq:F} reflect the power of isothermal coordinates in two dimensions: the free energy looks \emph{as if} the theory is formulated on flat space, with the curved geometry entering as an azimuthal vector potential $a_z = i\frac{p}{2}\partial\varphi$. In the case of a cone, this term corresponds to a magnetic monopole at the apex.
	
It is convenient to define a real valued dual variable $\Phi$ to the phase-field $\gamma$ of $Q$ such that
\begin{subequations}
	\begin{align}
		\partial\Phi &= -\frac{2i}{p}\partial \gamma \\
		\bar\partial\Phi &= \frac{2i}{p}\bar\partial \gamma,
	\end{align}
	\label{eq:PhiDefn}
\end{subequations}
in terms of which the free energy (Eq.~\eqref{eq:F}) becomes
\beq 
    \mathcal F = \frac{p^2}{4}J\int d^2z\left|\partial\left(\Phi - \varphi\right)\right|^2. 
\eeq
	
Note that in this paper, we freeze the geometry, which fixes the geometric potential $\varphi$. The geometry then determines the ground state configuration of $\gamma$, or equivalently, its dual $\Phi$. Upon suppressing the frozen $\varphi$ dependent part, we write $\mathcal F$ as
\beq 
    \mathcal F = \mathcal F_1 + \mathcal F_2 , \label{eq:FSum}
\eeq
where
\begin{align}
	\mathcal F_1 &=\frac{p^2}{4}J\int d^2z|\partial \Phi|^2 \\
	\mathcal F_2 &=-\frac{p^2}{4}J\int (\bar\partial\varphi \partial\Phi + \partial\varphi \bar\partial\Phi). \label{eq:F2}
\end{align}
Here, $\mathcal F_1$ is the elastic energy and $\mathcal F_2$ is the interaction energy between the $p$-atic texture and the geometry. We will see shortly that in analogy to electrostatics, $\Phi$ can be viewed as the electrostatic potential, sourced by topological defects. Using this idea, we show that $\mathcal F_1$ can be computed using the standard Green's function techniques, and $\mathcal F_2$ can be computed exactly through integration by parts via evaluating $\varphi$ at the topological defects, multiplied by the topological defect charges. Although it appears that $\mathcal F_1$ does not depend on $\varphi$, we show in Sec.~\ref{sec:F} that there is a subtle dependence on $\varphi$ coming from the short distance physics embodied in the defect core energies. 
	
Before we evaluate $\mathcal F$, we first review the description of topological defects using isothermal coordinates and then compute $\Phi$ in the presence of these singularities. For a $p$-atic, for a closed loop around a topological defect of charge $\sigma \in \mathbb{Z}/p$, $\gamma$ will wind by $2\pi p \sigma$. Moreover, by minimization of the free energy, $\gamma$ satisfies (away from the defects)
\beq 
    \partial\bar\partial \gamma = \frac{1}{4}\left(\p{}{x^2} + \p{}{y^2}\right) = 0 \label{eq:LaplaceEq} .
\eeq
Note that in Eq.~\eqref{eq:LaplaceEq}, there continues to be no $\varphi$ dependence--thus, the local physics is as if we are in flat space.
	
Near a defect at $z_j$, we have
\beq 
    \gamma \approx -\frac{ip}{2}\sigma_j \ln \frac{z - z_j}{\bar z - \overline{z_j}},
\eeq
which manifestly solves Eq.~\eqref{eq:LaplaceEq} and has the correct winding number. It follows that the dual variable $\Phi$ (defined by Eq.~\eqref{eq:PhiDefn}) satisfies
\beq 
    \partial\bar\partial \Phi = -\pi\sum_i\sigma_i\delta^2(z - z_i),\label{eq:Poisson}
\eeq
where defect $i$ is at position $z_i$ with charge $\sigma_i$. In other words, topological defects can be viewed as sources for $\Phi$, which, in analogy to electrostatics, behaves as the electric potential. In particular, as expected, the standard Green's function $G(z_1,z_2)$ is given by $\Phi(z_1;z_2)$, where the charge $\sigma$ is placed at $z_2$, i.e.,
\beq 
    G(z_1,z_2) = -\frac{1}{4\pi\sigma}\Phi(z_1;z_2). \label{eq:Green}
\eeq
	
Note that $\mathcal F_2$, upon integration by parts, can be written as
\beq 
    \mathcal F_2 = \frac{p^2}{4}J\int d^2z 2\Phi \partial\bar\partial\varphi = - \frac{p^2}{2}J\int d^2z \Phi R_{z\bar z}, 
\eeq
where $R_{z \bar z} = -\partial\bar\partial\varphi$ is the Ricci curvature. This implies that curvature gives rise to an effective two-dimensional charge density $\rho$, given by
\beq 
    \rho = -\frac{R_{z\bar z}}{2\pi}.
    \eeq
Thus, in general, regions of positive (negative) curvature give rise to negative (positive) charge density~\cite{park1996topological,bowick2009two}.
	
Now, to completely solve Eq.~\eqref{eq:LaplaceEq}, we must specify boundary conditions. We consider two different geometries: one without a boundary, such as the infinite plane or sphere, and the other with a boundary, such as the disk or cone.
	
\subsection{Plane/sphere}
	
To set the stage for disks and cones, we first consider planes and spheres. Taking into account the winding due to topological defects, the multi-defect solution to Eq.~\eqref{eq:LaplaceEq} that gives real values of $\gamma=p\theta$ is given by (see for example Ref.~\cite{vafa2020multi-defect})
\beq 
    \gamma = p\theta = -\frac{i}{2}\sum_i p\sigma_i[\ln(z - z_i) - \ln(\bar z - \bar z_i)], \label{eq:theta}
\eeq
where $z_i$ is the position of defect $i$ and $\sigma_i \in \mathbb Z/p$ is its charge. Thus, $Q$, $\Phi$, and $G(z_1,z_2)$ are given by
\begin{align}
	Q = Q^{z\ldots z} &= e^{-\frac{p\varphi}{2}} \prod_i \left(\frac{z - z_i}{|z - z_i|}\right)^{p\sigma_i}\label{eq:Q} \\
	\Phi &= -\sum_i \sigma_i \ln|z - z_i|^2\\
	G(z_1,z_2) &= \frac{1}{4\pi}\ln|z_1 - z_2|^2 . \label{eq:GreenPlane}
\end{align}
	
Note that although the geometric potential $\varphi$ that enters $Q$ differs for planes and spheres, the expression for $G(z_1,z_2)$ is the same for the plane and the sphere, independent of $\varphi$.
	
\subsection{Disk/cone}
	
We next consider the case of a geometry with a boundary, in particular a flat disk. We now must specify a boundary condition. In Ref.~\cite{zhang2022fractional}, free boundary conditions were imposed. Here, we consider tangential boundary conditions, by which we mean that $\nabla\gamma$ (the gradient of the order parameter phase) is tangential to the circumference of the base of the cone. We will assume that the total winding is $2\pi$, i.e., the total charge of the topological defects inside the cone is 1. For elementary defects,  the solution to Eq.~\eqref{eq:LaplaceEq} with tangential boundary conditions for $p$ defects each with charge $+1/p$ at $z_j$ is
\beq 
    \gamma = p\theta = -\frac{i}{2}\sum_{j=1}^{p}p\sigma_j\left[\ln\left( \frac{z - z_j}{\bar z - \overline{z_j}} \right)+ \ln\left( \frac{z - \tilde z_j}{\bar z - \overline{\tilde z_i}}  \right)\right],\label{eq:thetaCone}
\eeq
where the $\tilde z_j = R^2 /\overline{z_j}$ are the positions of (like-signed) image charges needed to impose the tangential boundary condition at $z = Re^{i\phi}$, with $\phi$ being the azimuthal angle and $R$ being the maximum radius in our isothermal coordinate system. (Recall that we encode the curved geometry through the geometric potential $\varphi$ and denote the azimuthal angle by $\phi$). As shown in Appendix~\ref{app:BC}, like-signed image charges leads from Eq.~\ref{eq:thetaCone} to a phase angle $\gamma(z)$ that is equal to the azimuthal angle $\phi(z)$ when $r=R$, independent of the location of the defect charges. The dual variable $\Phi$, (by a suitable choice of the integration constant), using equations~\eqref{eq:PhiDefn} and \eqref{eq:thetaCone}, is given by
\begin{align}
	\Phi &=  -\sum_{j=1}^{p}\sigma_j\left[\ln \frac{\left|z - z_j\right|^2}{R^2} + \ln \left|1 - \frac{z}{\tilde z_j}\right|^2\right] \nonumber\\
	&= -\sum_{j=1}^{p}\sigma_j\left[\ln \frac{\left|z - z_j\right|^2}{R^2} + \ln \left|1 - \frac{z\overline{z_j}}{R^2}\right|^2\right], \label{eq:PhiBC}
\end{align}
and hence the Green's function is
\beq 
    G(z_1,z_2) = \frac{1}{4\pi} \left[\ln \frac{\left|z_1 - z_2\right|^2}{R^2} + \ln \left|1 - \frac{z_1\overline{z_2}}{R^2}\right|^2\right].\label{eq:GreenBC}
\eeq
	
We would like to emphasize that the metric does not appear here so that $G(z_1,z_2)$ is the same for the disk and cone with tangential boundary conditions, although it does appear in $\mathcal F_2$ (Eq.~\eqref{eq:F2}).
	
\section{Evaluation of $\mathcal F$}
\label{sec:F}
	
We are now ready to evaluate the free energy $\mathcal F$. Integrating Eq.~\eqref{eq:FSum} by parts and using equations \eqref{eq:PhiBC} and \eqref{eq:GreenBC} lead to
\begin{align}
	\mathcal F_1 &= -(\pi p)^2J\sum_{mn} \sigma_m\sigma_n G(z_m,z_n)\label{eq:F1}\\
	\mathcal F_2 &= -2\pi\frac{p^2}{4}J\sum_m \sigma_m\varphi(z_m).
\end{align}
In analogy to electrostatics, we learn from $\mathcal F_2$ that $\varphi$ behaves as an additional contribution to the electrostatic potential due to the surface geometry. In particular, upon substitution of $\varphi = -\chi \ln (z\bar z)$ into $\mathcal F_2$, there is clearly an attraction (repulsion) of the positive (negative) defects to (from) the cone apex that is linear in the charge $\sigma_j$.
	
Naively, it appears that there is no $\varphi$ dependence in $\mathcal F_1$. However, there is a subtlety in evaluating Eq.~\eqref{eq:F1} due to the self-energy term coming from the $m=n$ terms in the double sum, pointed out in Ref.~\cite{vitelli2004anomalous}, which we now examine.
	
\subsection{Self-energy}
	
The self-energy is formally infinite, but this ignores the defect core size $\delta$, which sets a natural UV cut-off. Therefore, what we really mean by $G(z_m,z_m)$ is
\beq 
    G(z_m, z_m) = \lim_{d(z_m,z_n) \to \delta} G(z_n,z_m) ,
\eeq
where $d(z_m,z_n)$ is the distance between $z_m$ and $z_n$ and $\delta$ is the minimum distance determined by hard core repulsion between liquid crystal molecules. By definition of the metric,
\beq 
    d(z_m,z_n) = e^{\varphi(z_m)/2}|z_m - z_n| = \delta 
\eeq
and thus
\beq|z_m - z_n| = \delta e^{-\varphi(z_m)/2}.\label{eq:d}
\eeq
Now, using the fact that for small point separation $z_m$ and $z_n$, the singular part of $G(z_m,z_n)\sim \frac{1}{4\pi} \ln |z_m - z_n|^2$, we can write
\beq 
    G(z_m,z_n) \sim \frac{1}{4\pi} \ln |z_m - z_n|^2 + \widehat G(z_m,z_m), \label{eq:GFormal}
\eeq
where $\widehat G(z_m,z_m)$ is non-singular at short distances. Upon substitution of Eq.~\eqref{eq:d} into Eq.~\eqref{eq:GFormal}, we get
\beq 
    G(z_m,z_m) = \frac{1}{4\pi}\left(-\varphi(z_m) + \ln\delta^2\right) + \widehat G(z_m,z_m). \label{eq:GResidual}
\eeq
For example, for the plane or sphere, Eq.~\eqref{eq:GreenPlane} gives $\widehat G(z_m,z_m) = 0$. For a flat plane, $\varphi = 0$ as well, but for the unit sphere, $\varphi = 2\ln \frac{2}{1+|z|^2}$, thus contributing to Eq.~\eqref{eq:GResidual}.  For the disk/cone geometry Eq.~\eqref{eq:GreenBC} gives
\beq 
    \widehat G(z_m,z_m) = \frac{1}{4\pi} \ln \left|1 - \frac{|z_m|^2}{R^2}\right|^2 .
\eeq
Thus, $\mathcal F$ (after dropping the constant term involving $\delta$ in $\mathcal F_1$) becomes
\begin{align}
	\mathcal F &= -(\pi p)^2 J \sum_{mn} \sigma_m\sigma_n G(z_m,z_n) \nonumber \\
	&\qquad -\frac{\pi p^2}{2} J \sum_m \left(\sigma_m - \frac{1}{2}\sigma_m^2\right)\varphi(z_m) \label{eq:masterF}\\
	&=-\pi\frac{p^2}{2} J\left\{\sum_{m<n}\sigma_m\sigma_n\left[\ln \frac{|z_m - z_n|^2}{R^2} + \ln \left|1 - \frac{z_m\overline{z_n}}{R^2}\right|^2 \right] \right. 
	\nonumber \\
	&\qquad \qquad + \sum_j\sigma_j^2\ln\left(1 - \frac{|z_j|^2}{R^2}\right)  
	\nonumber\\
	&\qquad \qquad \left. -\chi \sum_j \left(\sigma_j - \frac{\sigma_j^2}{2}\right)\ln \frac{|z_j|^2}{R^2} \right\} ,
\end{align}
where we are now denoting $\widehat G(z_m,z_m)$ as $G(z_m, z_m)$ in the double sum. The first equality holds in general, and the second equality is specialized to the case of a cone with deficit angle $2\pi\chi$.
	
In other words, the self-energy term gives rise to $\sigma_m^2 \varphi/2$, which represents an additional contribution to the geometric interaction, and depends quadratically on the defect charge $\sigma_m$, in agreement with the general results of Refs~\cite{park1996topological} and \cite{vitelli2004anomalous}. We now provide some intuition for the $\sigma_m^2$ term by deriving the self-energy term explicitly in the context of a cone.
	
\subsubsection{Self-energy on a cone}
	
\begin{figure}
	\centering
	\includegraphics[width=\columnwidth]{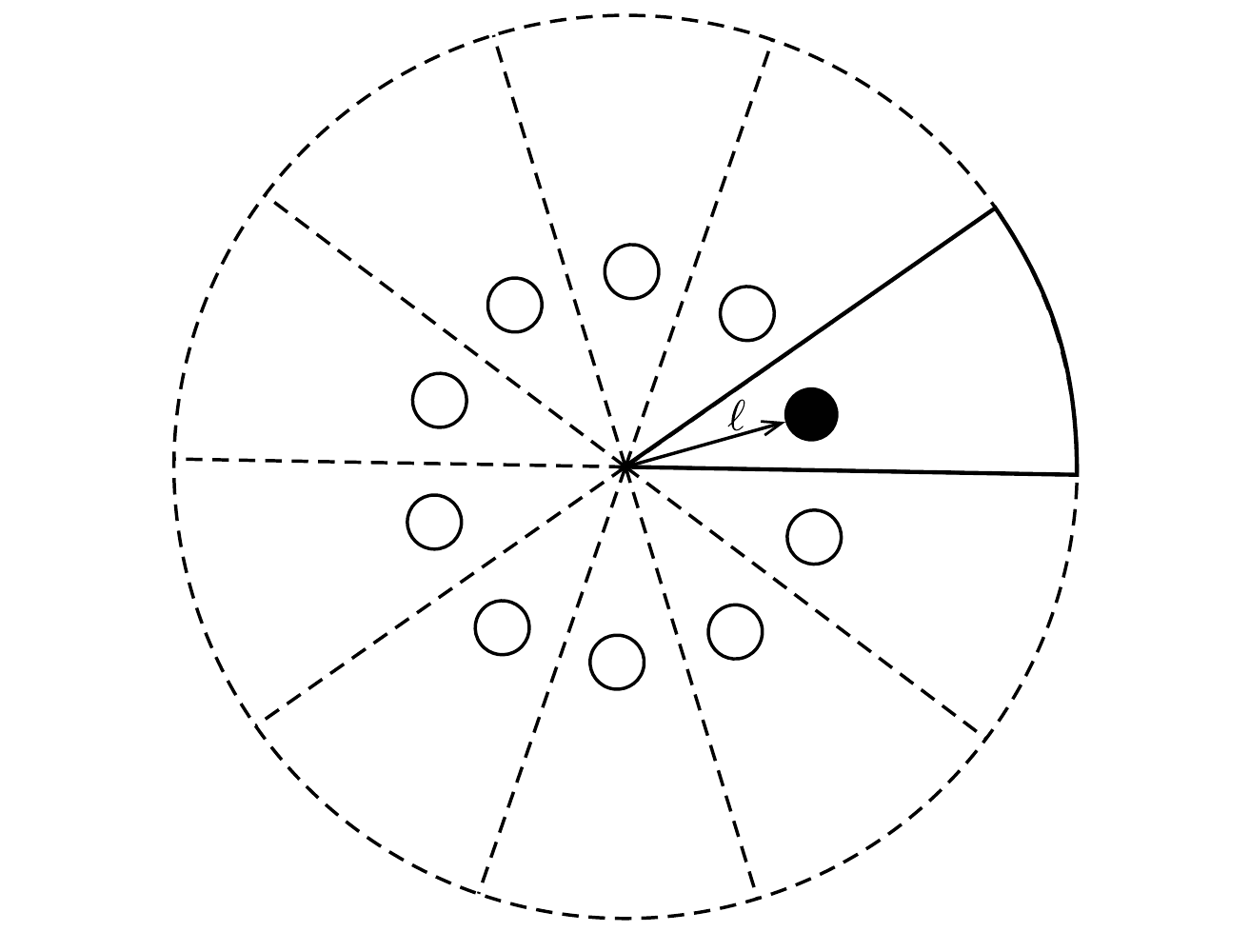}
	\caption{Schematic of $\tilde z$ geometry for $\sin\beta = 1/n$, for $n=10$. The conical singularity is represented by the star at the origin, the topological defect is represented by the black dot in the black wedge, and $n-1$ image charges are represented by white circles in the dashed wedges.}
	\label{fig:images}
\end{figure}
	
Here we first explain the quadratic dependence of the self-energy on the defect charge and see intuitively why it is repulsive for a cone. Let's consider a cone with special commensurate half-angle $\beta$ such that $\sin\beta = 1/n$ for a given integer $n$. As shown in Fig.~\ref{fig:images}, such a cone is equivalent to $\mathbb{R}^2/\mathbb{Z}_n$. What this means is that if we have a topological defect, then because of the $\mathbb{Z}_n$ it is as if there are 1 physical and $n-1$ image charges, a charge in each of the $n$ wedges (see Fig.~\ref{fig:images} for a schematic). Then, it's clear that the interaction between a defect and the geometric defect charge of the cone is quadratic in the topological charge and also repulsive.
	
We'll now make this argument more quantitative. Let $\ell$ denote the distance between any defect and the origin, and let $r$ and $s$ denote defects, where $r,s = 0,\ldots,n-1$. Then the interaction energy between a pair of defects $r$ and $s$ on a plane, (upon substituting Eq.~\eqref{eq:GreenPlane} into Eq.~\eqref{eq:F1}), is  $\mathcal F_1 =-\pi\frac{p^2}{2}J\sigma_m^2 \ln d_{rs}^2$, where $d_{rs} = \ell|e^{2\pi i (r/n)} - e^{2\pi i (s/n)}|$ is the distance between the defects. Since $d_{rs} \propto \ell$ for all pairs, the total elastic energy is given by
\begin{align}
	E &=  -\pi \frac{p^2}{2}J\frac{1}{n}{n\choose 2} \sigma_m^2 \ln \ell^2 + \const \nonumber \\
	&= -\pi\frac{p^2}{4} J \sigma_m^2(n-1)  \ln \ell^2 + \const
\end{align}
The factor of $1/n$ is due to the fact that the physical space is one of these $n$ wedges, and the binomial factor ${n\choose 2}$ counts all of the pair-wise interactions. Now, on using the following coordinate transformation (Eq.~\eqref{eq:ztilde})
\beq 
    \ell = \frac{|z_m|^{1-\chi}}{1-\chi}
\eeq
the energy is, up to a constant,
\beq 
    E =   -\pi \frac{p^2}{4} J (n-1)(1-\chi) \sigma_m^2 \ln |z_m|^2 .
\eeq
Using $1 - \chi = 1/n$ (for the special case $\sin\beta = 1/n$) then leads to
\beq 
    E =  -\pi \frac{p^2}{4} J\sigma_m^2 \chi \ln |z_m|^2 = \pi \frac{p^2}{4} J \sigma_m^2 \varphi(z_m),
\eeq
recovering for a cone the quadratic term in Eq.~\eqref{eq:masterF}.	
	
\section{Ground states of defects on disk and cone}	
\label{sec:groundStates}

Here we compute the ground state defect configuration for the disk and cone with tangential boundary conditions. Tangential boundary conditions provide a much richer arena than the free boundary conditions of Ref.~\cite{zhang2022fractional}, because defects on the cone flanks can be an intrinsic part of the ground state. For a cone, substituting $\varphi = -\chi \ln z \bar z$ and Eq.~\eqref{eq:GreenBC} into Eq.~\eqref{eq:masterF} immediately gives (with $\chi = 1-\sin\beta$)
	
\begin{align}
	\mathcal F &= -\pi\frac{p^2}{2} J\left\{\sum_{m<n}\sigma_m\sigma_n\left[\ln \frac{|z_m - z_n|^2}{R^2} + \ln \left|1 - \frac{z_m\overline{z_n}}{R^2}\right|^2 \right] \right. \nonumber\\
	&\qquad \left. + \sum_j\sigma_j^2\ln\left(1 - \frac{|z_j|^2}{R^2}\right) -\chi \sum_j \left(\sigma_j - \frac{\sigma_j^2}{2}\right)\ln \frac{|z_j|^2}{R^2} \right\}\label{eq:FCone}.
\end{align}
We interpret each term in turn. The first term (the double sum) is the usual elastic interaction between pairs of defects, including image charges. The second term is the self-energy, which would need to be added to any microscopic defect core energy $E_\mathrm{c}$. The final term represents the interaction between a topological defect and the geometry~\cite{vitelli2004anomalous}, specialized to the cone. Note that the cone apex develops an effective topological charge of $-\chi$. This is also compatible with the recent results of \cite{zhang2022fractional} in finding the ground state configuration of a $p$-atic liquid crystal on a cone with free boundary conditions, which is equivalent to minimizing the magnitude of the effective charge at the cone apex. In particular, the minimum energy configuration considered in~\cite{zhang2022fractional} can have some number $s_0$ of charge $+1/p$ defects at the cone apex absorbed from the free outer rim. On keeping the $|\partial \varphi|^2$ term in Eq.~\ref{eq:F}, and converting to physical coordinates using Eq.~\ref{eq:ztilde}, we obtain a ground state free energy of $(\pi J p^2/2(1-\chi)) q_\mathrm{eff}^2\ln \tilde R/a$, where $q_\mathrm{eff} = -\chi + s_0/p$ is the effective charge at the cone apex, $s_0 = \underset{s}{\mathrm{argmin}} |-\chi + s/p|$ is the number of defect charges  that optimally screens out the geometric contribution $-\chi$, and $\tilde R$ the longitudinal length of the cone along the flanks. The result is consistent with Ref.~\cite{zhang2022fractional}.
	
Moreover, a topological defect of charge $\sigma_j$, when interacting with the cone apex, behaves as if it had an effective charge
\beq 
    Q_\mathrm{eff} = \sigma_j - \sigma_j^2/2. 
\eeq
Hence an elementary positive (negative) defect with charge $\sigma_j = \pm 1/p$ will be attracted to (repelled from) the cone tip. (We note in passing that these attractions and repulsions will be reversed for hyperbolic cones, e.g., the surfaces formed when negative disclinations are allowed to relax into the third dimension~\cite{seung1988defects}.)
	
The general strategy for constructing ground states is that topological defects (including possible image charges) interact with each other via a $2d$ logarithmic Coulombic interaction, i.e., same-sign defects want to be as far away from each other as possible. Since the cone apex has a negative effective topological charge, depending on the deficit angle, it will absorb as many positive defects it can until the net charge at the apex becomes positive, in which case no additional defects will be absorbed. The remaining defects will then be as far away as possible from the cone apex. It seems plausible that they would lie equally spaced on a ring, a conjecture confirmed by our numerical simulations. We will now describe this picture more quantitatively.
	
\subsection{Disk}
	
To set the stage for a cone, we first consider $p$-atics on disks with tangential boundary conditions. In this case, setting $\chi=0$ in Eq.~\eqref{eq:FCone} reduces to
\begin{align}
	\mathcal F &= -\pi\frac{p^2}{2} J\left\{\sum_{m<n}\sigma_m\sigma_n\left[\ln \frac{|z_m - z_n|^2}{R^2} + \ln \left|1 - \frac{z_m\overline{z_n}}{R^2}\right|^2 \right] \right. \nonumber \\
	& \qquad \qquad \left. + \sum_j\sigma_j^2\ln\left(1 - \frac{|z_j|^2}{R^2}\right)  \right\}.
\end{align}
	
We have suppressed a contribution to the core energy of the defects, usually modeled by a term $E_c \sum_j \sigma_j^2$. This term prefers elementary defects of minimal charge $\sigma = \pm 1/p$, since $1/p^2 + 1/p^2 < (2/p)^2$, which motivates us to consider only elementary defects in this paper.
	 
For $p$ defects each of charge $\sigma_j = +1/p$ equally spaced on a concentric ring of radius $d=xR$ in the isothermal coordinates, i.e., $z_j = d e^{2\pi i (j/p)}$, $j = 0,\ldots,p-1$, the free energy is computed to be
\beq 
    \mathcal F = -\pi\frac{p^2}{2} J \left[\frac{1}{p^2}\frac{p(p-1)}{2}\ln x^2 + \frac{p}{p^2}\ln(1 - x^{2p})\right] + \const .\label{eq:FDisk}
\eeq
In deriving Eq.~\eqref{eq:FDisk}, we used the fact that for $\sigma_j = +1/p$ and $z_j = de^{2\pi i (j/p)}$, we have 
\begin{align}        
    &\sum_{m<n}\sigma_m\sigma_n\ln \left|1 - \frac{z_m\overline{z_n}}{R^2}\right|^2 + \sum_j\sigma_j^2\ln\left(1 - \frac{|z_j|^2}{R^2}\right) \nonumber\\
	& \qquad = p \sum_{j=0}^{p-1}\frac{1}{p^2}\ln \left(1 - x^2 e^{2\pi i (j/p)}\right) \nonumber\\
	& \qquad = \frac{p}{p^2} \ln\prod_{j=0}^{p-1}\left(1 - x^2 e^{2\pi i (j/p)}\right) \nonumber\\
	& \qquad = \frac{p}{p^2} \ln\left(1 - x^{2p}\right)
	.
\end{align}
Minimizing Eq.~\eqref{eq:FDisk} over the dimensionless flank distance $x$ gives
\beq 
    \label{eq:dxR_disk} x =  \left(\frac{p-1}{3p-1}\right)^{\frac{1}{2p}} . 
\eeq
	
\subsection{Cone}
	
We now return to the generalized case of the cone and consider the following defect configuration: $k$ defects of charge $+1/p$ equally spaced on a ring at a distance $d=xR$ on the cone flank, i.e., for these defects, $z_j = d e^{2\pi i (j/p)}$, $j=0,\ldots,k-1$, and the remaining $p-k$ defects at the cone apex. Then the free energy becomes (up to a constant)
\beq 
    \mathcal F = -\pi\frac{p^2}{2} J \left[\frac{1}{p^2}\frac{k(k-1)}{2}\ln x^2 + \frac{k}{p^2}\ln(1 - x^{2k}) + k\frac{\chi'}{p}\ln x^2\right],
\eeq
where 
\beq 
    \chi' = - \left(1 - \frac{1}{2p}\right)\chi + \frac{p-k}{p} .
\eeq
The $\chi'$ term determines whether a defect is absorbed by the core. These transitions happen at critical cone angles such that
\beq 
    \chi_c'=0 \implies \chi_c = \frac{2(p-k)}{2p-1} ,\label{eq:chi_c}
\eeq
and $\mathcal F$ here is minimized when 
\beq 
    x = \left(\frac{k - 1 +2p\chi'}{3 k - 1 +2p \chi'}\right)^{\frac{1}{2k}}. \label{eq:dxR}
\eeq
On using Eqs.~\eqref{eq:ztilde} and \eqref{eq:chi}, the fractional distance $\tilde x$ along the flank (for the unrolled  coordinates in Fig.~\ref{fig:coordinates}b) is
\beq 
    \tilde x = \left(\frac{k - 1 +2p\chi'}{3 k - 1 +2p \chi'}\right)^{\frac{\sin\beta}{2k}}.\label{eq:xtilde}
\eeq
Note that here $k$ is chosen such that
\beq  
    \chi' - 1/p < 0 \le \chi' ,
\eeq
or equivalently,
\beq 
    \frac{2(p-k-1)}{2p-1} \le \chi < \frac{2(p-k)}{2p-1} .
\eeq
	
In other words, there are three general cases for ground state configurations:
\begin{enumerate}
	\item $\frac{2(p-1)}{2p-1}<\chi$:  the ground state will consist of $p$ defects of charge $+1/p$ that have been swallowed up by the cone apex.
	\item $\frac{2(p-k-1)}{2p-1} \le \chi < \frac{2(p-k)}{2p-1}$: the ground state will consist of $p-k$ defects of charge $1/p$ at the apex and $k$ defects at $z_j = d e^{2\pi i (j/k)}$, $j = 0, 1, \dots, k-1$ ($d = xR$ is determined in Eq.~\eqref{eq:dxR} by minimizing the free energy).
	\item $\chi<0$: the ground state will consist of $p$ defects of charge $1/p$ at $z_j = d e^{2\pi i (j/p)}$, $j = 0,\ldots,p-1$ ($d$ is determined by minimizing the free energy).
\end{enumerate}
	
\begin{figure}
	\centering
	\includegraphics[width=\columnwidth]{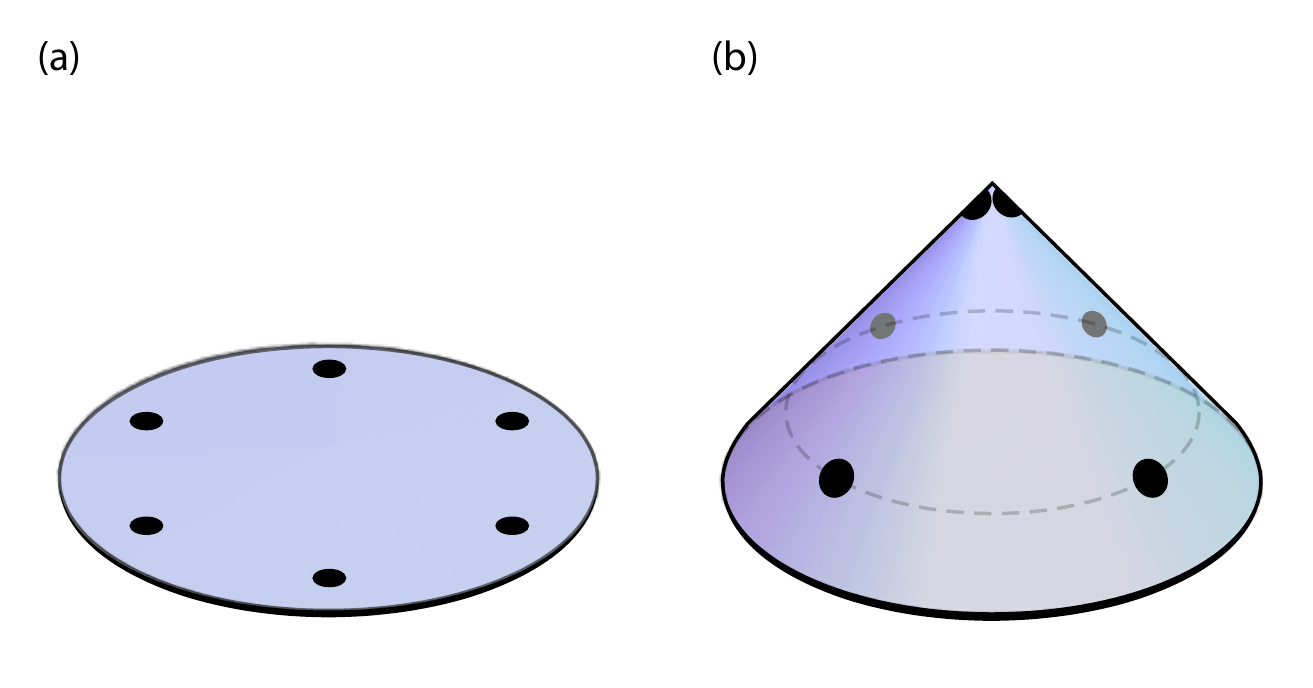}
	\caption{Schematic illustration of ground state defect configurations on a disk and a cone. The positive Gaussian curvature at the cone apex gives rise to a geometrical background charge that attracts like-signed defects in the $p$-atic liquid crystal. (a): A hexatic ($p=6$) liquid crystal on a flat disk has six defects with charge $+1/6$ distributed evenly along the angular direction at positions given by Eq.~\eqref{eq:dxR_disk} with $p=6$. (b): A cone with angle $\sin\beta = 4/6$ absorbs two of those defects onto the apex, leaving four defects on the flanks, again evenly distributed along the azimuthal direction. Defects are depicted by black dots.}
	\label{fig:3}
\end{figure}

To summarize, we expect that the ground state of a $p$-atic on a flat disk with tangential boundary conditions at $r = R$ has $p$ defects of charge $+1/p$ spaced out evenly on a concentric ring at distance $d = xR$ (Eq.~\eqref{eq:dxR_disk}) from the disk center (see Fig.~\ref{fig:3}a for $p=6$). As the cone angle increases (the surface deviates more from flatness), the cone apex absorbs the $+1/p$ defects one by one at certain values of $\chi$, while the rest of the defects lie equally spaced along a ring at some distance $d(\chi)$ that depends on the cone angle (see Eq.~\eqref{eq:dxR}) from the apex (see Fig.~\ref{fig:3}b, which illustrates $p=6$ and $\chi = 1/3$.). 
	
\subsection{Maier-Saupe model and numerics}
	
\begin{figure*}
    \centering
    \includegraphics{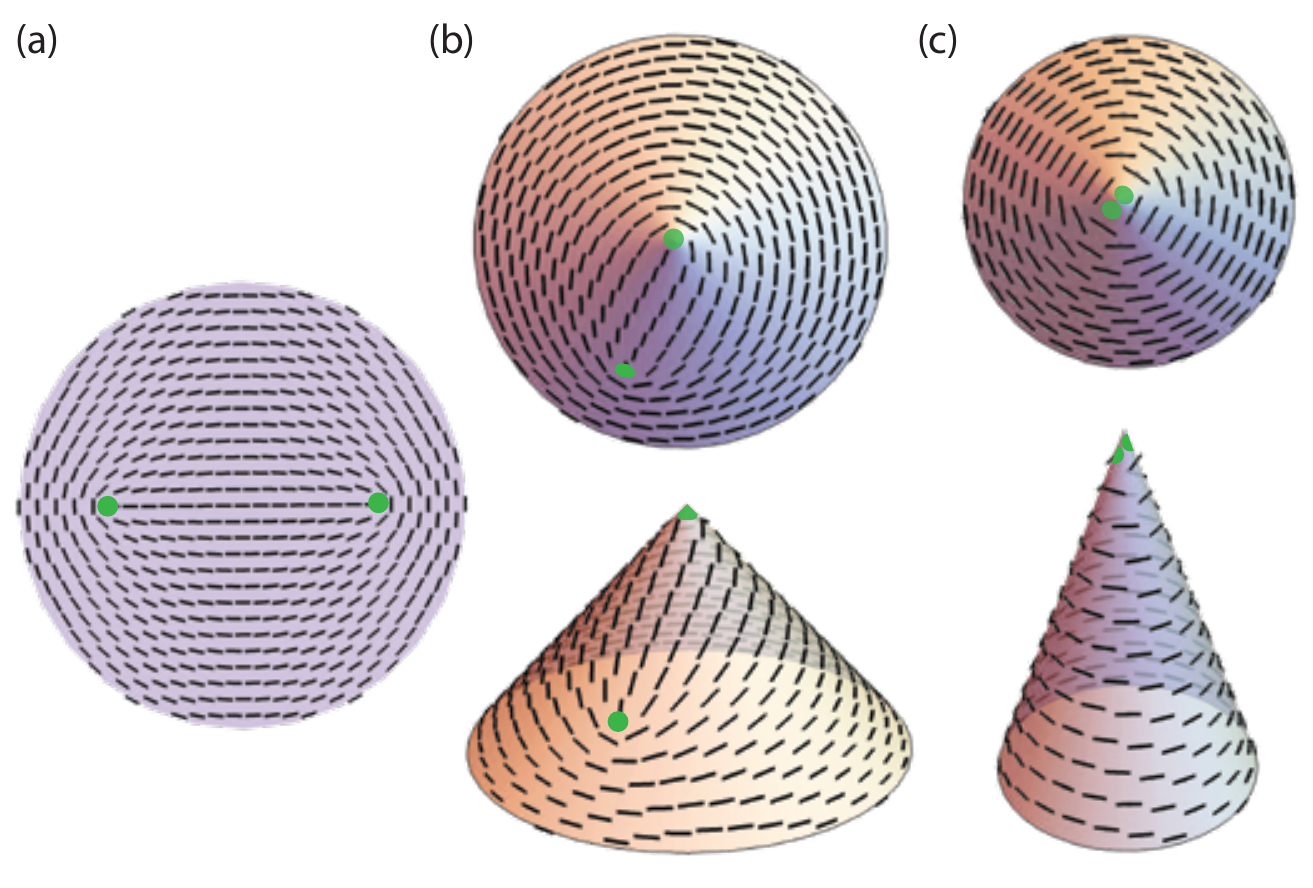}
    \caption{Ground state numerical textures for a nematic ($p=2$) liquid crystal with tangential boundary conditions for various values of $\chi=1-\sin\beta$, where $\beta$ is the cone half-angle. (a) On a flat disk ($\chi = 0$), there are two $+1/2$ defects, labeled with green dots, at positions given by Eq.~\eqref{eq:dxR_disk} with $p=2$. (b) On the surface of a cone corresponding to $\chi = 1/3$, there is one $+1/2$ defect on the flank and another at the cone apex. (c) On a cone corresponding to $\chi = 2/3$, there are two $+1/2$ defects at the cone apex, leaving none on the flanks. For (b) and (c), we show both top and perspective views of the cone.}
    \label{fig:tex}
\end{figure*}
We now check our continuum results above with ground state energy minimizations on lattices. The discrete Hamiltonian follows from the Maier-Saupe model for a two-dimensional system of $p$-atic liquid crystals on curved surfaces, with interactions that align nearest neighbors~\cite{selinger2015introduction},
\begin{align} 
    H &= -J' \sum_{\langle i j \rangle}[T_p(\hat m_i \cdot \hat m_j)] \nonumber \\
	&= -J' \sum_{\langle i j \rangle} \left[ \cos (p (\theta_i - \theta_j + A_{ij})) - 1 \right],
    \label{eq:H_micro_cone}
\end{align}
where $i,j$ are site indices, $\langle ij \rangle$ indicates nearest neighbors, $\hat m_i$ is an orientational unit vector attached to a liquid crystal molecule at site $i$, $\theta_i$ is the orientation angle of molecule $i$ in the local frame of site $i$, and $A_{ij}$ is the rotation angle induced by parallel transport between site $i$ and $j$. $T_p(x)$ is the $p$-th Chebychev polynomial~\cite{zwillinger2018crc}, and $J'$ is the microscopic Maier-Saupe coupling strength between molecules at neighboring sites. As shown in Appendix~\ref{app:J}, $J'$ maps onto the coarse-grained parameters in our free energy as $ J' = J/4$ for a square lattice and $ J' = J/4 \sqrt{3}$ for a triangular lattice. On the surface of a cone, the vectors describing the orientation of $p$-fold symmetric molecules need to be parallel transported to the local frame of its neighbor before their dot product is taken. As shown in Eq.~\eqref{eq:H_micro_cone}, the interaction energy between two neighboring molecules at sites $i$ and $j$ is hence modified by a rotation angle $A_{ij}$ that the molecule undergoes during the parallel transport. 
	
Using the interaction energy in Eq.~\eqref{eq:H_micro_cone} and fixing the orientation vectors $\hat m_i$ at the base of the cone to obey tangential boundary conditions, we simulate $p$-atic liquid crystals on lattices on the surfaces of cones using the Python Broyden-Fletcher-Goldfarb-Shanno (BFGS) algorithm~\cite{broyden1970convergence,fletcher1970new,goldfarb1970family,shanno1970conditioning}. Our numerical energy minimizations focus on the cone angles for which a regular triangular or square mesh is especially straightforward to generate~\cite{zhang2022fractional}. The numerical ground state textures for a nematic liquid crystal on a disk $\chi = 0$ and cones corresponding to $\chi = 1/3$ and $2/3$ are shown in Fig.~\ref{fig:tex}. The total apex defect charge for all commensurate cones simulated are tabulated in Table~\ref{tab:apex}. 
	
Note that vectors at the cone apex do not have a well defined orientation, since the azimuthal coordinate $\phi$ is undefined there. We thus perform all energy minimizations with the orientation vector at the apex removed. 
\begin{table}[]
\centering
$$\begin{array}{c|c|c|c|c|c|c}
\hline \chi & p=1 & p=2 & p=3 & p=4 & p=5 & p=6 \\
\hline 0 & 1 & 0 & 0 & 0 & 0 & 0 \\
\hline \frac{1}{6} & 1 & \frac{1}{2} & \frac{1}{3} & \frac{1}{4} & \frac{1}{5} & \frac{1}{6} \\
\hline \frac{3}{4} & 1 & \frac{1}{2} & \frac{1}{3} & \frac{1}{4} & \frac{2}{5} & \frac{2}{6} \\
\hline \frac{4}{6} & 1 & \frac{1}{2} & \frac{1}{3} & \frac{2}{4} & \frac{2}{5} & \frac{2}{6} \\
\hline \frac{3}{6} & 1 & \frac{1}{2} & \frac{2}{3} & \frac{2}{4} & \frac{3}{5} & \frac{3}{6} \\
\hline \frac{2}{6} & 1 & \frac{2}{2} & \frac{2}{3} & \frac{3}{4} & \frac{3}{5} & \frac{4}{6} \\
\hline \frac{1}{4} & 1 & \frac{2}{2} & \frac{2}{3} & \frac{3}{4} & \frac{4}{5} & \frac{5}{6} \\
\hline \frac{1}{6} & 1 & \frac{2}{2} & \frac{3}{3} & \frac{3}{4} & \frac{4}{5} & \frac{5}{6} \\
\hline
\end{array}$$
\caption{Apex defect charges extracted from numerical energy minimizations of $p$-atics on commensurate cone angles.}
\label{tab:apex}
\end{table}
	
\begin{figure*}
	\centering
	\includegraphics[width=\textwidth]{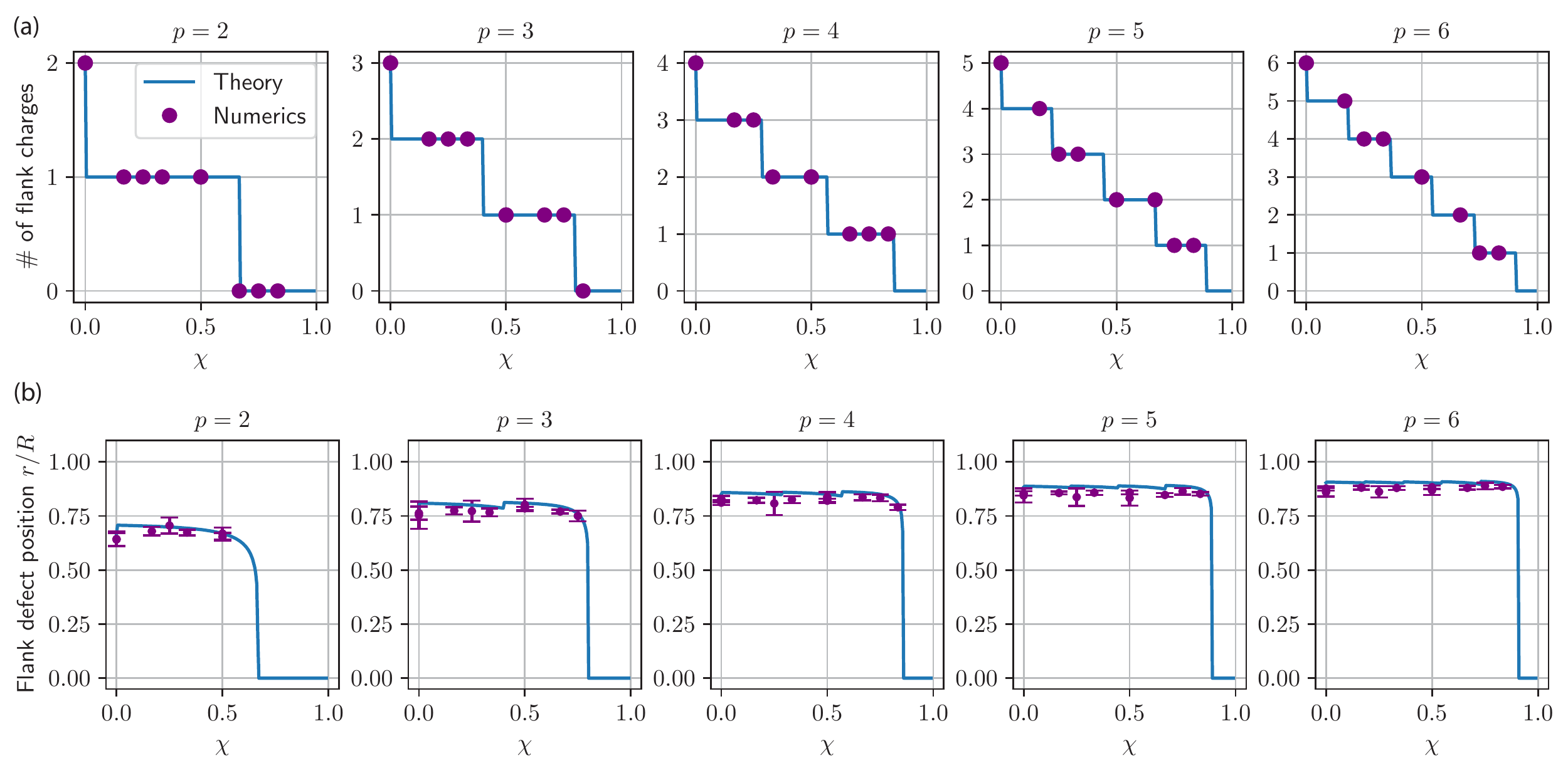}
	\caption{
	Top row: plots of number of flank charges as a function of $\chi=1-\sin\beta$, where $\beta$ is the cone half-angle. Purple markers are from numerical energy minimization, and blue line is theoretical prediction for defect absorption transitions (Eq.~\eqref{eq:chi_c}). Bottom row: plots of flank defect positions in the ground states of $p$-atics on cones as a function of $\chi$. Purple markers are from numerical energy minimization, and blue curve is theoretical prediction (Eq.~\eqref{eq:xtilde}).}
	\label{fig:4}
\end{figure*}

Table~\ref{tab:apex} summarizes our numerical finding for the defect content of the apex for $\sin\beta = 1 - \chi$ = 1/6, 1/4, 2/6, 3/6 = 2/4, 4/6, 3/4 and 5/6. Fig.~\ref{fig:4} shows excellent agreement between theory and numerics on both the total number of flank charges and their radial position as a function of $\chi$. See Appendix~\ref{app:ground} for a complete summary of all defect configurations in the ground state we have explored numerically.
	
\section{Conclusion}
\label{sec:conclusion}
	
Our simplified model coupling $p$-atic liquid crystal order to geometry based on isothermal coordinates reveals that the cone apex develops an effective topological charge proportional to the deficit angle of the cone. This observation leads to a mechanism of defect absorption and emission at the cone apex with one important conclusion about ground state configurations: compared to the defect configuration on a disk, positive (negative) defects are absorbed (emitted) by the cone apex, with transitions and positions of the flank defects intricately depending on the deficit angle and the charges of the defects.
	
To connect to biological systems, we must include non-equilibrium effects, such as activity. Recently, a tensorial hydrodynamic theory of $p$-atics was investigated on flat surfaces~\cite{giomi2021hydrodynamic,giomi2021longranged}. In the presence of activity, a motile nematic $+1/2$ defect would interact with the cone apex depending on its position and polarization relative to the azimuthal direction of the cone, which could lead to interesting orbits in the absence of noise. For example, it is conceivable that a nematic $+1/2$ defect could slingshot around the cone apex on a trajectory approximating a geodesic as if under the influence of gravity due to the negative effective charge of the apex. It would be interesting to study the dynamics of active topological defects on curved surfaces.
	
It would also be worth exploring defect configurations on cones with free boundary conditions at finite temperatures. Entropic effects might cause the cone apex to cough up some of the defects it has swallowed in increasing temperatures.
	
\begin{figure}
    \centering
    \includegraphics[width=\columnwidth]{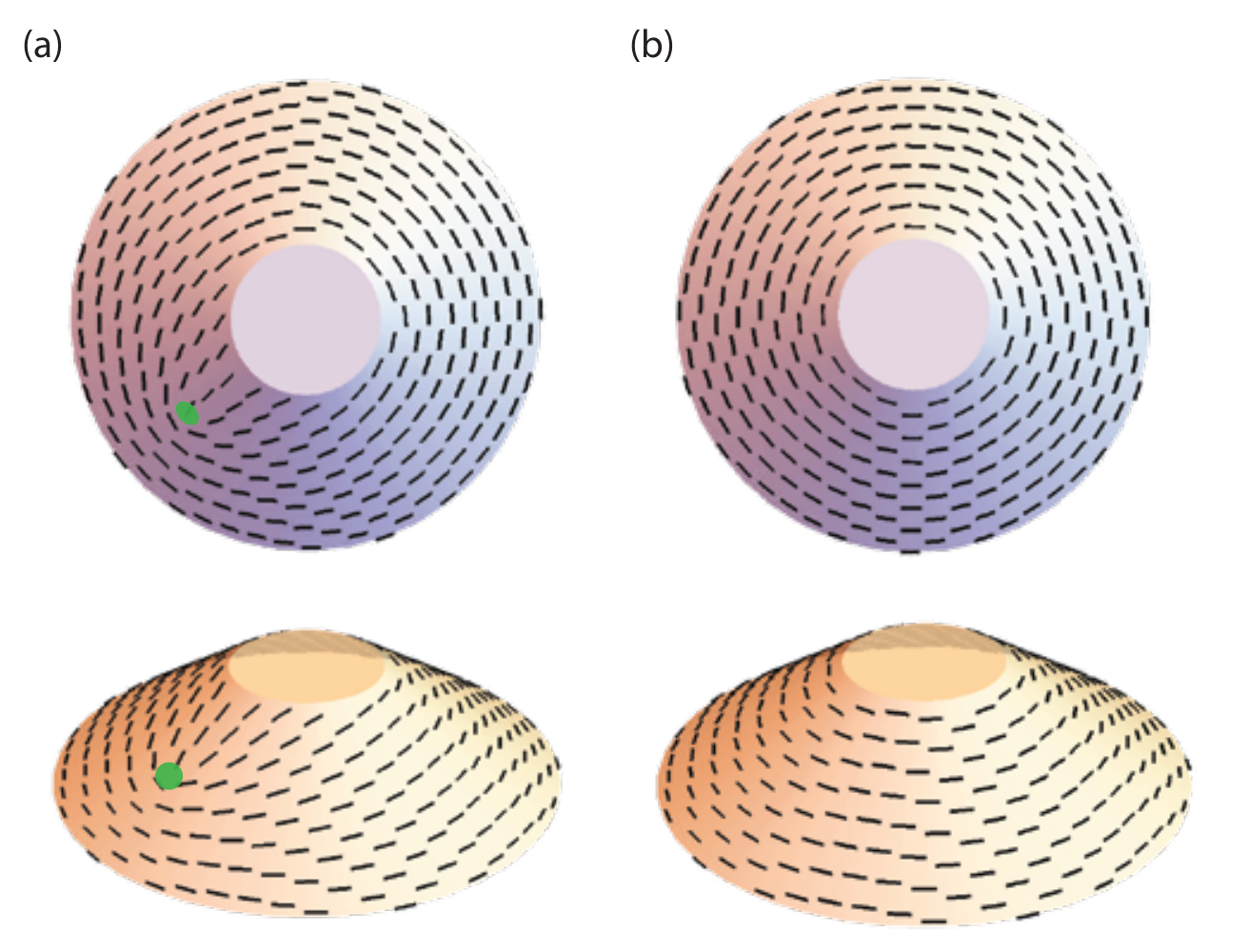}
    \caption{Nematic textures from numerical energy minimizations of $p=2$ liquid crystals on truncated $\chi = 1/6$ cones with an inner truncation of $r_\mathrm{inner} = 3$ lattice constants and $r_\mathrm{outer} = 10$, with (a) tangential BC at the bottom rim and free BC at the top rim (b) tangential BC at both the top and bottom rims. Green circles indicate $\sigma = +1/2$ defects.
    }
	\label{fig:trunc_BC}
\end{figure}

It is also interesting to consider variants of the boundary conditions considered here. The topological nature of the geometrical frustration associated with the cone makes it clear that slightly truncated cones would behave in a similar fashion, provided we maintain tangential boundary conditions at the base and impose free boundary conditions at the top. In our numerical minimizations, we removed a single site at the cone tip, which is a limiting example of free boundary conditions at the apex. This point is illustrated by Fig.~\ref{fig:trunc_BC}a below, which shows both perspective and rolled out views of a $p = 2$ conical texture with inner rolled out radius $r_\mathrm{inner} = 3$ and $r_\mathrm{outer} = 10$ lattice constants an a cone angle such that $\chi = 1/6$. Both the texture and the position of the single $\sigma = +1/2$ defect on the cone flank are essentially indistinguishable from the defect we find with only a single apex site removed.
	
On the other hand, imposing tangential boundary conditions at both the top and bottom of a truncated cone does change the ground state. As one might expect, there are no defects on flanks, and the frozen $p$-atic texture simply interpolates between the tangential boundary conditions at the top and bottom (See Fig.~\ref{fig:trunc_BC}b). The case of tangential boundary conditions at the top of a truncated cone and free boundary conditions at the bottom is also interesting. We leave a full understanding of this intriguing problem for general $p$ and arbitrary cone angles to a future investigation.  
	
\begin{figure}
    \centering
    \includegraphics[width=\columnwidth]{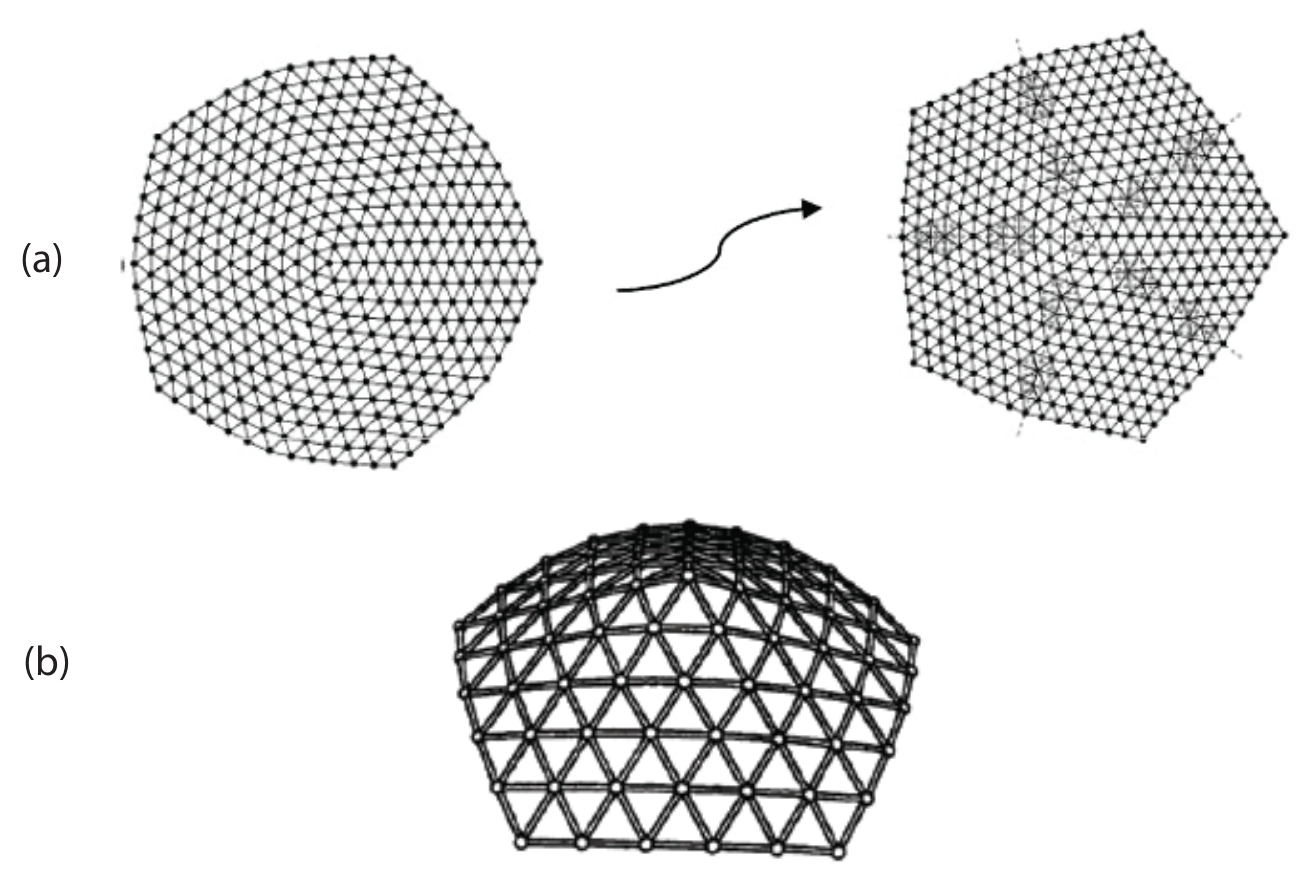}
    \caption{(a) A flat two-dimensional crystal with a five-fold disclination at the origin can lower its energy by forming five grain boundaries to screen the central disclination charge. (b) If allowed to buckle into the third dimension, the crystal with the disclination can lower its energy further without needing to form any grain boundaries. Adapted from Refs.~\cite{moore1999absence,seung1988defects}.}
    \label{fig:GB}
\end{figure}
	
Finally, we comment briefly on the challenging problem of determining the ground states of, say, triangular crystals on cones with arbitrary opening angles. It is natural to expect grain boundaries, like the grain boundary scars discussed for spheres in Ref.~\cite{bowick2000interacting}, in the ground state.  In the simple disk-like example shown in Fig.~\ref{fig:GB}a, there is a rotation of a hexatic order parameter in the crystal of 72 degrees around the rim, (somewhat similar to the 360 degree rotation caused by tangential boundary conditions applied to a crystal spanning an annulus, considered in Ref.~\cite{bruinsma1982motion}.)   Without grain boundaries in the disk, the energy on the left grows like $YR^2$, where $Y$ is the young's modulus and $R$ is the radius. However, introducing dislocations will lower this energy.   The highly anisotropic interactions between dislocations on the right leads to 5 grain boundaries with 12 degree jumps in crystal orientation, and produces an energy which grows linearly in $R$ and hence is preferred, at least in flat space~\cite{moore1999absence}. In both cases, there is a five-fold disclination at the ``apex'' of the disk.
	
However, if this disclination is put on a cone with just the right cone angle, like this one with $\chi = 1/6$~\cite{seung1988defects}, all grain boundaries vanish (see Fig.~\ref{fig:GB}b), and the energy will be lowered even more, to now depend logarithmically on the system radius $R$. Less pointy cones should produce intermediate numbers of grain boundaries, somewhat similar to the variable number of flank defects we have found for liquid crystal order on cones with tangential BC on the rim. We expect similar configurations when the boundary conditions enforces a 360 degree rotation at the edge and the cone angle is varied.   
	
\acknowledgments{It is a pleasure to acknowledge helpful conversations with Paul Hanakata, Suraj Shankar, Abigail Plummer, Rudro Biswas, and Alberto Fernandez-Nieves. This work is partially supported by the Center for Mathematical Sciences and Applications at Harvard University (F. V.). G.H.Z. acknowledges support by the National Science Foundation Graduate Research Fellowship under Grant No. DGE1745303. This work was also supported by the NSF through the Harvard Materials Science and Engineering Center, via Grant No. DMR-2011754 (D. R. N.).
	
\appendix

\section{Positive and negative image charges on the cone}
\label{app:BC}

In this appendix, we illustrate the utility of isothermal coordinates by exploring the boundary conditions associated with both positive and negative image charges for a $p$-atic liquid crystal on a cone with $p$ elementary defects, each with minimal charge $\sigma_j = + 1/p$. The local angle of the $p$-atic order parameter is given by a simple generalization of Eq.~\eqref{eq:thetaCone}, 
\beq
    \theta(z, \bar z) = \frac{\gamma (z, \bar z)}{p} = - \frac{i}{2} \sum_j \sigma_j \left[ \ln \left ( \frac{z - z_j}{\bar z - \bar z_j} \right) \pm \ln \left( \frac{z - \tilde z_j}{\bar z - \overline {\tilde z_j}} \right) \right],
\eeq
where an equal number of image charges with charges $\pm 1/p$ are located at $\tilde z_j = R^2/\bar z$, and the $\pm$ signs correspond to positive and negative image charges, respectively. Note that the denominators in the logarithms ensure that the phase angles are real. It is straightforward to check that an isolated defect at position $z_j$ causes $\theta(z, \bar z)$ to rotate by $ 2 \pi / p$ on a small contour surrounding the defect. 

We first show that positive image charges indeed reflect the tangential boundary conditions associated with a $p$-atic that rotates uniformly by $2 \pi$ around the edge at $z = R e^{i \phi}$, where $R$ is the radius of the base of the cone, independent of the location of the defects. We first use
\begin{eqnarray}
    \partial_z = \frac{1}{2} \left( \frac{ \partial}{\partial x} - i \frac{ \partial}{\partial y} \right), \quad \partial_{\bar z}  = \frac{1}{2} \left( \frac{ \partial}{\partial x} + i \frac{ \partial}{\partial y} \right)
\end{eqnarray}
to evaluate the quantity 
\beq
    \hat z \cdot ( \vec r \times \vec \nabla) \theta(x,y) = x \partial_y\theta - y \partial_x \theta = \frac{1}{i} ( \bar z \partial_{\bar z} - z \partial_z) \theta(z, \bar z),
\eeq
where $\vec r = (x,y) = r (\cos \phi, \sin \phi)$ and we shall eventually set $r = R$. It is straightforward to show that 
\beq \label{eq:bound}
    \hat z \cdot (\vec r \times \vec \nabla) \theta = \sum_{j=1}^p \sigma_j \left[ \mathrm{Re} \left( \frac{z}{\bar z - \bar z_j} \right) + \mathrm{Re} \left( \frac{z}{\bar z - \overline {\tilde z_j}} \right) \right],
\eeq
where $\tilde z_j = R^2/\bar z_j$.
Upon setting $z = Re^{i \phi},~z_j = r_j e^{i \phi_j}$, and $\tilde z_j = (R^2/r_j) e^{i \phi_j}$, we find that
\begin{eqnarray}
    \mathrm{Re} \left( \frac{z}{z - z_j} \right) = \frac{1 - \frac{r_j}{R} \cos(\phi_j - \phi)}{1 + \left( \frac{r_j}{R} \right)^2 - \frac{2 r_j}{R} \cos(\phi_j - \phi)}
\end{eqnarray}
and
\begin{eqnarray}
    \mathrm{Re} \left( \frac{z}{z - \tilde z_j} \right) = \frac{\left( \frac{r_j}{R} \right)^2 - \frac{r_j}{R} \cos(\phi_j - \phi)}{1 + \left( \frac{r_j}{R} \right)^2 - \frac{2 r_j}{R} \cos(\phi_j - \phi)}. 
\end{eqnarray}
Upon inserting these results into Eq.~\ref{eq:bound}, we see immediately that
\begin{eqnarray}
    \hat z \cdot (\vec r \times \vec \nabla) \theta \Big |_{r = R} = \sum_{j=1}^p \sigma_j = 1 ,
\end{eqnarray}
\textit{independent} of coordinate $\phi$ on the rim of the base of the cone and of the locations $\{ z_j = r_j e^{i \phi_j} \}$ of $p$ defect charges on the cone. Thus, the orientation of the $p$-atic molecules, even in the presence of defects, rotate uniformly at the rim. For the problem considered in this paper, with $p$ positive defects each with charge $\sigma_j = 1/p$ on the cone, we have
\begin{eqnarray} \label{eq:tg_b}
    \oint_{r = R} \vec \nabla \theta \cdot d \vec \ell = 2 \pi \sum_{j=1}^p \sigma_j = 2 \pi, 
\end{eqnarray}
which is a manifestation of Gauss' law.

For the case of boundary conditions provided by \textit{negative} image charges, a very similar calculation shows that
\begin{align}
    \vec r \cdot \vec \nabla \theta \Big|_{r = R} &= \left( x \frac{ \partial}{\partial x} + y \frac{ \partial}{\partial y} \right) \theta \Big |_{r = R} \nonumber\\
    &= (\bar z \partial_{\bar z} + z \partial_z) \theta(z,\bar z) \Big |_{z = Re^{i \phi}} \nonumber \\
    &= 0.
    \label{eq:f_b}
\end{align}
With these negative image charge boundary conditions, the radial component of the phase gradient vanishes, so that the phase gradient is again tangential. However, the tangential component of the gradient now varies in a complicated fashion as a function of the azimuthal position along the boundary. Indeed, it is readily shown that
\beq
    \hat z \cdot (\vec r \times \vec \nabla) \theta \Big |_{r = R} = \sum_{j=1}^p \sigma_j \left( \frac{R^2 - r^2_j}{R^2 + r^2_j - 2 r_j R \cos(\phi_j - \phi)} \right).
\eeq
Despite this complicated azimuthal variation, one can show that the integral of the phase gradient along the rim still results in this simple form, identical to the first equality of Eq.~\ref{eq:tg_b},
\begin{eqnarray}
    \oint_{r = R} \vec \nabla \theta \cdot d \vec \ell = 2 \pi \sum_{j=1}^p \sigma_j.
\end{eqnarray}
We must now decide on the value of $\sum_{j=1}^p \sigma_j$, under these more complex negative image charge boundary conditions. In the ground state, we expect this quantity to vanish, because any defects in the interior of the cone would be attracted to and annihilate with their oppositely-signed image charges outside the cone, as shown for a planar boundary in Fig.~\ref{fig:def_ann}.
\begin{figure}
    \centering
    \includegraphics[width=\columnwidth]{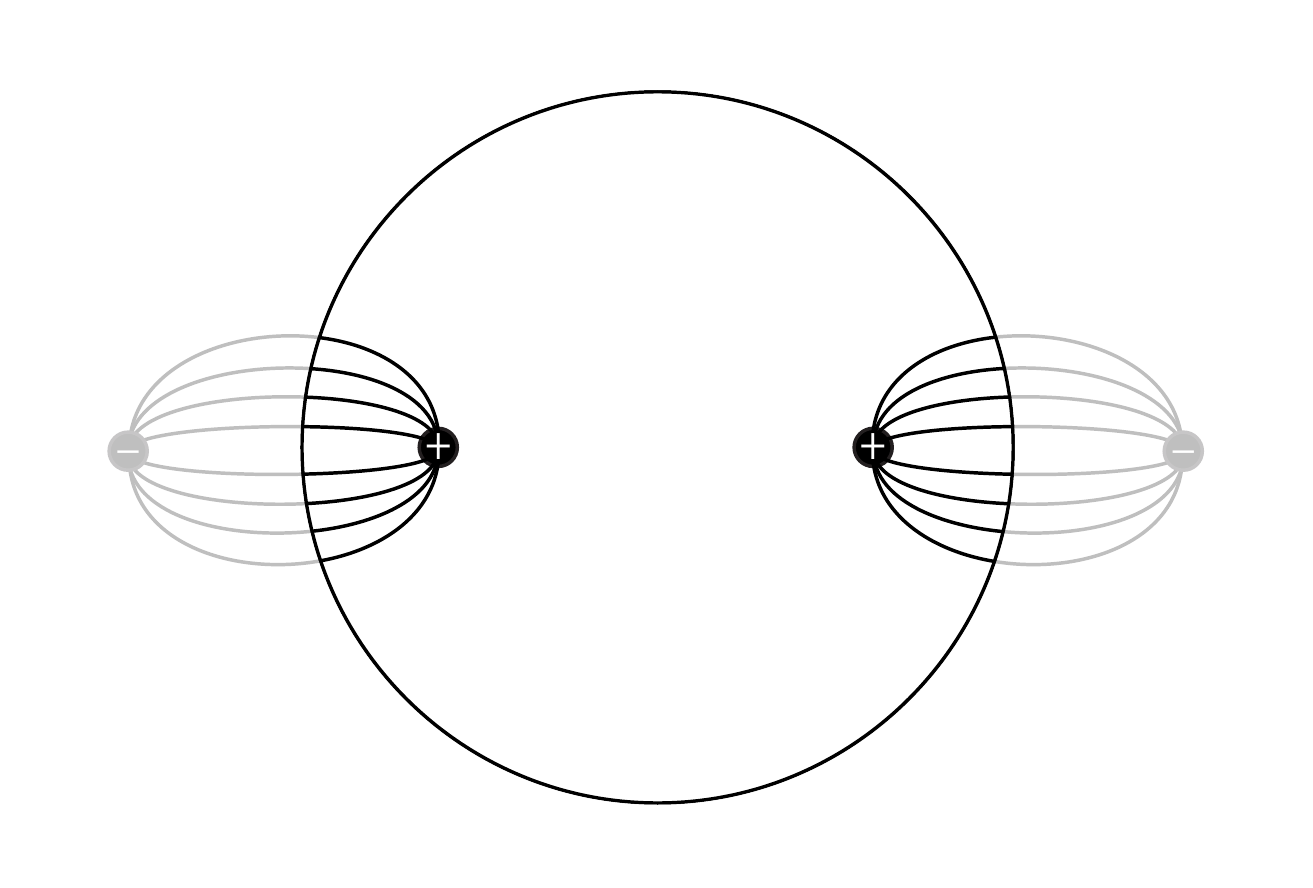}
    \caption{Electric field lines (without arrows) of two positive charges inside a disk and their negative image charges outside the disk.}
    \label{fig:def_ann}
\end{figure}
A possible exception is defects at the cone apex, which is allowed because then the image charge would then be infinitely far away.  This absence of defects in the interior of the cone in the ground state is consistent with the results of Ref.~\cite{zhang2022fractional} for free boundary conditions. In this case, the free energy is minimized when enough defects are added at the apex (denoted by $s_0$ in Eq.~53 of Ref.~\cite{zhang2022fractional}) such that the magnitude of the effective charge $q_\mathrm{eff}$ at the apex (including the geometric contribution) is minimal, i.e. $q_\mathrm{eff} = |-\chi + s_0/p|$ where $s_0 = \underset{s}{\mathrm{argmin}} |-\chi + s/p|$. It is appropriate to characterize the disks and cones studied in Ref.~\cite{zhang2022fractional} as having ``free boundary conditions,'' because the orientations of the $p$-atic molecules are unconstrained at the boundary, except by their neighbors in the tangential direction. Hence, the gradient of their phase angle will vanish normal to the boundary. This is indeed the case, as shown in Eq.~\ref{eq:f_b}. Note that the situation is quite different from tangential boundaries, as defects near boundaries are repelled by their images. 

\section{Relation between Maier-Saupe lattice coupling and $J$} \label{app:J}
	
\begin{figure}
    \centering
    \includegraphics[width=\columnwidth]{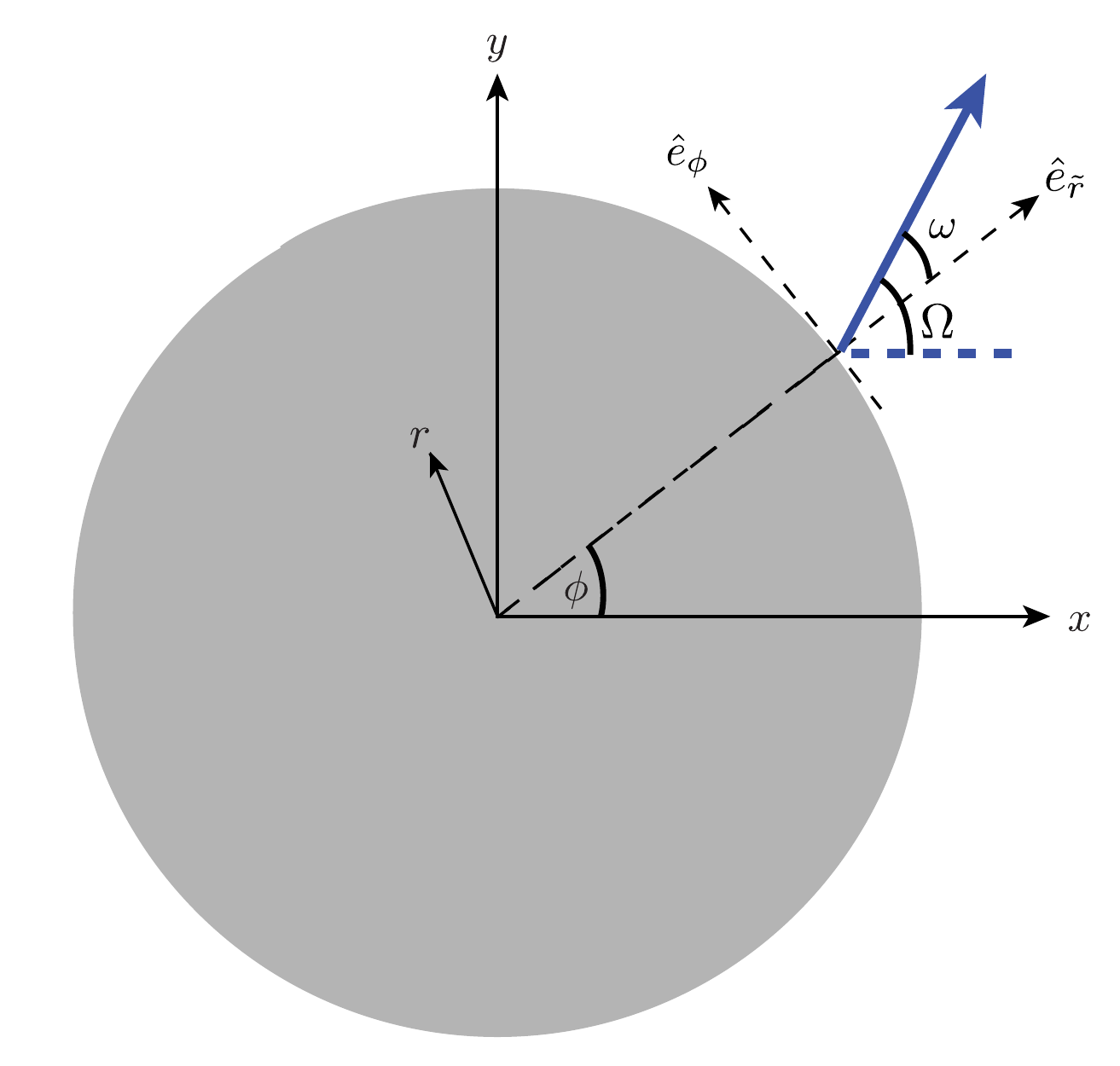}
    \caption{$\omega$ denotes the angle of the director field $\hat n$ of the liquid crystal molecule (blue arrow) relative to the local orthogonal axes on the cone surface. $\Omega$ indicates the angle of the director field $\hat n$ on the squashed isothermal cone, relative to the real axis of the 2d complex plane.}
    \label{fig:Omega}
\end{figure}
	
In this section, we clarify the relation between the continuum free energy in isothermal coordinates and the Maier-Saupe lattice Hamiltonian (Eq.~\eqref{eq:H_micro_cone}) used for simulations.
	
Let $\Omega$ denote the angle of the liquid crystal molecule on the isothermal cone relative to the real axis of the complex plane (see Fig.~\ref{fig:coordinates}a and \ref{fig:Omega}). We can write Eq.~\eqref{eq:F} in terms of the angle $\Omega$. For a $p$-atic, $\Omega$ is related to $\gamma$ (the angle of the $p$-atic tensor component $Q$) as (${\bf Q} \sim [\hat n^{\otimes p}]_{TS}$, where $M_{TS}$ indicates the traceless symmetric part of $M$, see for example Ref.~\cite{giomi2021hydrodynamic}),
\begin{eqnarray}
    p \Omega = \theta.
\end{eqnarray}
The free energy in Eq.~\eqref{eq:F} can then be written as,
\begin{eqnarray}\label{eq:F0_Omega}
    F_0 &=&  p^2 J \int dz d\bar z  \left |\partial \Omega - \frac{i}{2} \partial \varphi \right |^2,
\end{eqnarray}
with $J = K+K'$. Upon making the substitutions, which follow from the relations $r = \sqrt{z \bar z}$ and $\phi = \frac{1}{2 i} \ln (z/\bar z)$, and remembering that $\varphi(z,\bar z) = -\chi \ln(z \bar z)$, we have
\begin{align}
    \partial \Omega &= \frac{e^{-i\phi}}{2} \left( \partial_r \Omega - \frac{i}{r} \partial_\phi \Omega \right) \\
    \frac{i}{2} \partial \varphi &= - i \frac{e^{-i\phi}}{2} \frac{\chi}{r},
\end{align}
and Eq.~\eqref{eq:F0_Omega} becomes
\beq 
    F_0 =   \frac{p^2}{4} J \int_0^{2 \pi} d\phi \int_0^R dr r \left( \left |\partial_r \Omega \right|^2 + \left| \frac{ \partial_\phi \Omega}{r} - \frac{\chi}{r} \right|^2 \right). \label{eq:F02} 
\eeq
On using the following relations between cone coordinates (see Eq.~\eqref{eq:ztilde} and Fig.~\ref{fig:coordinates}),
\beq 
    rdr = [(1-\chi) \tilde r]^{\frac{2}{(1-\chi)} - 1} d\tilde r, \quad |\partial_r \Omega|^2 = [(1-\chi) \tilde r]^{2- \frac{2}{(1-\chi)}} |\partial_{\tilde r} \Omega|^2,
\eeq
we can rewrite the free energy in terms of the longitudinal coordinate $\tilde r$ of the conic surface as
\begin{align} 
    &F_0 = \frac{p^2}{4} J \int_0^{2 \pi} d\phi \nonumber\\
    &\quad \times\int_0^{\tilde R} d\tilde r (1-\chi) \tilde r  \left( \left |\partial_{\tilde r} \Omega \right|^2 + \left| \frac{ \partial_\phi \Omega}{(1-\chi) \tilde r} - \frac{1}{(1-\chi) \tilde r} + \frac{1}{\tilde r} \right|^2 \right) . 
\end{align}
Next, upon rewriting $\Omega$ in terms of the angle $\omega$ that the director field makes with respect to the $\hat e_{\tilde r}$ axis of the local frame,
\begin{eqnarray} \label{eq:omega_phi}
    \Omega = \omega + \phi,
\end{eqnarray}
we obtain
\beq
    \label{eq:F0_omega}
    F_0 =  \frac{p^2}{4} J\int_0^{2 \pi} d\phi \int_0^{\tilde R} d\tilde r (1-\chi) \tilde r \left( \left |\partial_{\tilde r} \omega \right|^2 + \left| \frac{ \partial_\phi \omega}{(1-\chi) \tilde r} + \frac{1}{\tilde r} \right|^2 \right) .
\eeq
Eq.~\eqref{eq:F0_omega} is precisely the continuum version of the Maier-Saupe lattice Hamiltonian in Eq.~\eqref{eq:H_micro_cone}, with $J' = \frac{1}{4} J$ for a square lattice and $J' = \frac{1}{4 \sqrt{3}} J$ for a triangular lattice~\cite{zhang2022fractional}. 

\section{Effect of truncation} \label{app:trunc}

\begin{figure}
    \centering
    \includegraphics[width=\columnwidth]{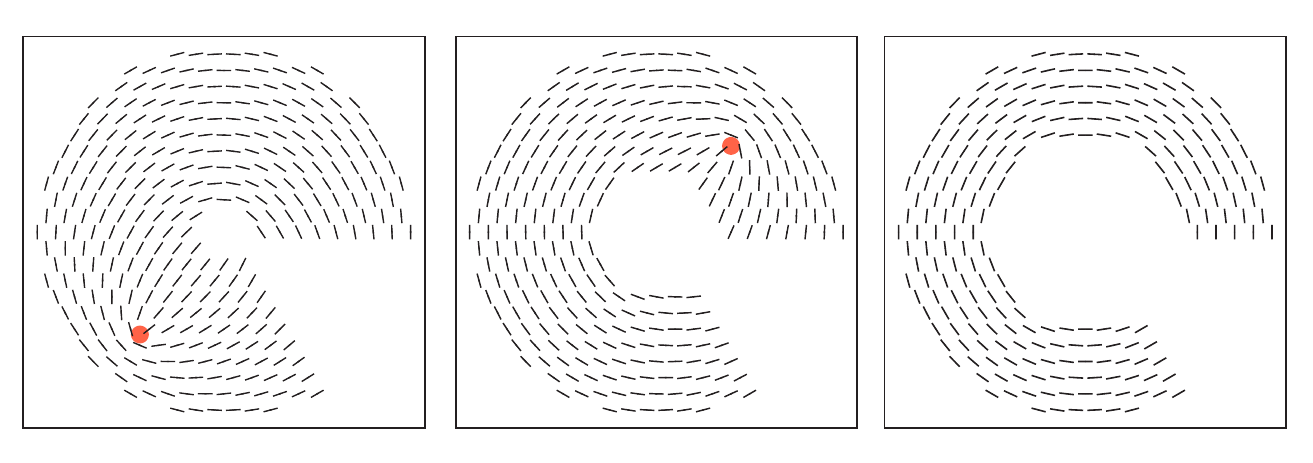}
    \caption{Defect configurations for a nematic liquid crystal with cone angle $\chi = 1/6$ and inner truncation radii $r_\mathrm{inner} =2,~4,~6$. When the truncation radius of the cone is sufficiently small, a truncated cone with free boundary conditions on the inner rim retains qualitatively the same distribution of defect charges on the flank and the apex as that for the untruncated cones studied in this paper. When the cone is truncated too sufficiently close to the outer rim, flank defects get absorbed to the center of the inner rim.}
    \label{fig:trunc}
\end{figure}

Here we consider the geometry of a truncated cone. In isothermal coordinates, without loss of generality, let the radius of the inner boundary be $R_1$ and the radius of the outer boundary be $R_2=1$, with $r = R_1/R_2 = R_1 < 1$. Using the method of images to impose free boundary conditions at the inner boundary ($z = r e^{i\phi}$) and tangential boundary conditions at the outer  boundary ($z = e^{i\phi}$) leads to the following modified Green's function
\begin{align}
    G(z_1, z_2) &= \frac{1}{4\pi}\left[\ln |z_1 - z_2|^2 \right. \nonumber\\
    &\quad - \sum_{n=0}^\infty (-1)^n \ln \left|z_1 - r^{2n+2}z_2\right|^2 \nonumber\\
    &\quad - \sum_{n=0}^\infty (-1)^n \ln \left|r^{2n+2}z_1 - z_2\right|^2 \nonumber \\
    &\quad + \sum_{n=0}^\infty (-1)^n \ln \left|z_1 \overline{z_2} - r^{-2n}\right|^2 \nonumber\\
    &\quad \left. - \sum_{n=0}^\infty (-1)^n \ln \left|z_1 \overline{z_2} - r^{2n+2}\right|^2 \right] .
\end{align}
In terms of the $q$-Pochhammer symbol,
\beq 
    (a;q) \equiv \prod_{n=0}^\infty(1 - a q^n),
\eeq
the Green's function can be expressed as
\begin{align}
    G(z_1,z_2) &= \frac{1}{4\pi}\left[\ln|z_1 - z_2|^2 - \ln \left|\frac{\left(r^2\frac{z_2}{z_1}; r^4\right)}{\left(r^4\frac{z_2}{z_1}; r^4\right)}\right|^2 - \ln \left|\frac{\left(r^2\frac{z_1}{z_2}; r^4\right)}{\left(r^4\frac{z_1}{z_2}; r^4\right)} \right|^2 \right. \nonumber \\
    &\qquad \left. + \ln \left|\frac{\left(z_1\overline{z_2}; r^4\right)}{\left(z_1\overline{z_2}r^2; r^4\right)}\right|^2 + \ln \left|\frac{\left(\frac{r^2}{z_1\overline{z_2}};r^4\right) }{\left(\frac{r^4}{z_1\overline{z_2}};r^4\right)} \right|^2\right],
\end{align}
or compactly as
\begin{align}
    &G(z_1,z_2) = \nonumber\\
    &\quad \frac{1}{4\pi}\ln \left|(z_1 - z_2)\frac{\left(r^4\frac{z_2}{z_1}; r^4\right)\left(r^4\frac{z_1}{z_2}; r^4\right)\left(z_1\overline{z_2}; r^4\right)\left(\frac{r^2}{z_1\overline{z_2}};r^4\right)}{\left(r^2\frac{z_2}{z_1}; r^4\right)\left(r^2\frac{z_1}{z_2}; r^4\right)\left(z_1\overline{z_2}r^2; r^4\right)\left(\frac{r^4}{z_1\overline{z_2}};r^4\right)}\right|^2.
\end{align}
Note that by using $(a;0) = 1-a$, it is easy to check that as $r\to0$, the original Green's function for the cone (Eq.~\eqref{eq:GreenBC}) is recovered.

The defect configurations we find in this work appear robust to small truncations of the cone top. Fig.~\ref{fig:trunc} shows the numerical ground state textures of a nematic liquid crystal on a cone with $\chi = 1/6$ and flank length $\tilde R = 10$. The number of flank defects stay the same for truncation radius $\tilde r_0 < 6$ and get absorbed to the center of the inner rim when $\tilde r_0 \geq 6$ approaches more than half way to the outer rim. In the latter limit, the effect of boundary condition starts to dominate that of geometry and the cone starts to behave more as a cylinder.

\section{Ground state textures} \label{app:ground}

The following pages summarize the results of our extensive numerical calculations of ground state configurations of $p$-atics on disks and cones with tangential boundary conditions at the base edges, obtained from numerical energy minimizations of the Hamiltonian in Eq.~\eqref{eq:H_micro_cone}. The configurations are arranged by row according to $\chi = 1 - \sin \beta$, where $\beta$ is the half cone angle, and by column according to liquid crystal symmetry parameter $p$. 
Additionally, although $\chi = 0/6,~3/6$ are equivalent in value to $\chi = 0/4,~2/4$, the corresponding numerical ground states are shown separately here, where $\chi = 0/6,~3/6$ indicate simulations done on a triangular lattice mesh, while $\chi = 0/4,~2/4$ indicate those done on a square lattice mesh. Minimal defects in the ground state of charge $+1/p$ on the disk and the cone flanks are marked with red circles. The defect configurations corresponding to cone angles that admit tilings with both square and triangular lattices are essentially indistinguishable.

\begin{figure*}
    \centering
    \includegraphics[width=1\textwidth]{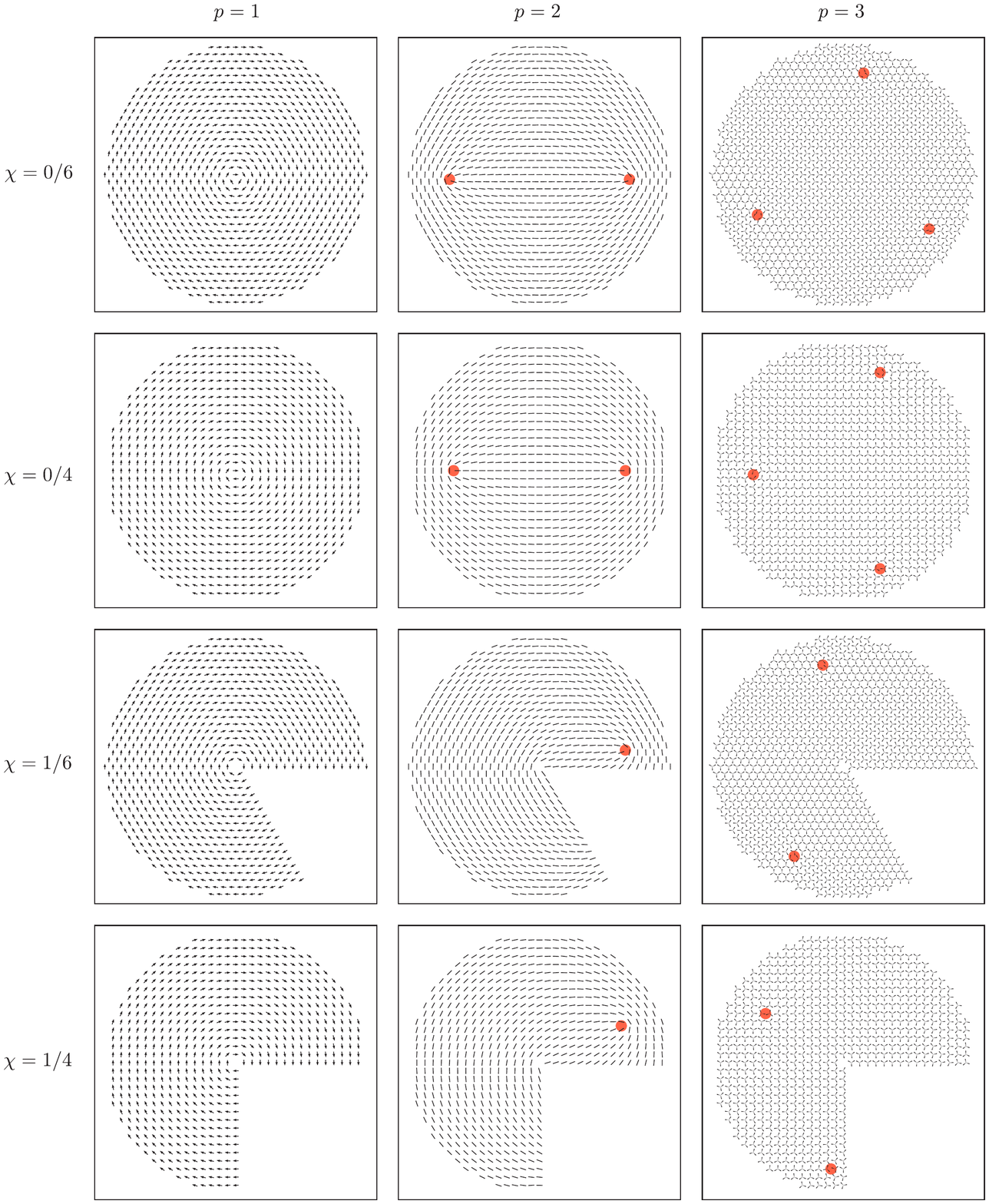}
\end{figure*}

\begin{figure*}
    \centering
    \includegraphics[width=1\textwidth]{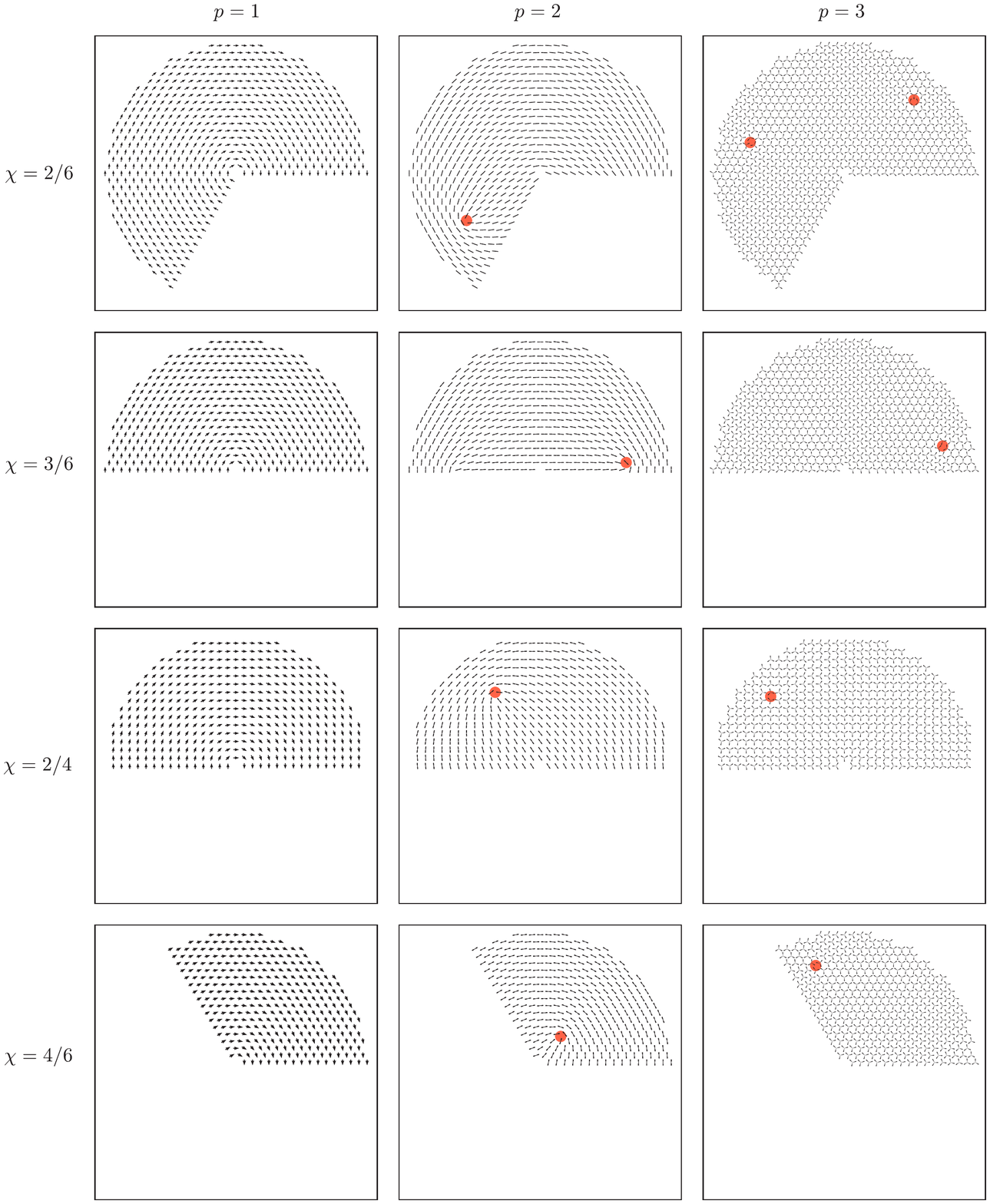}
\end{figure*}

\begin{figure*}
    \centering
    \includegraphics[width=1\textwidth]{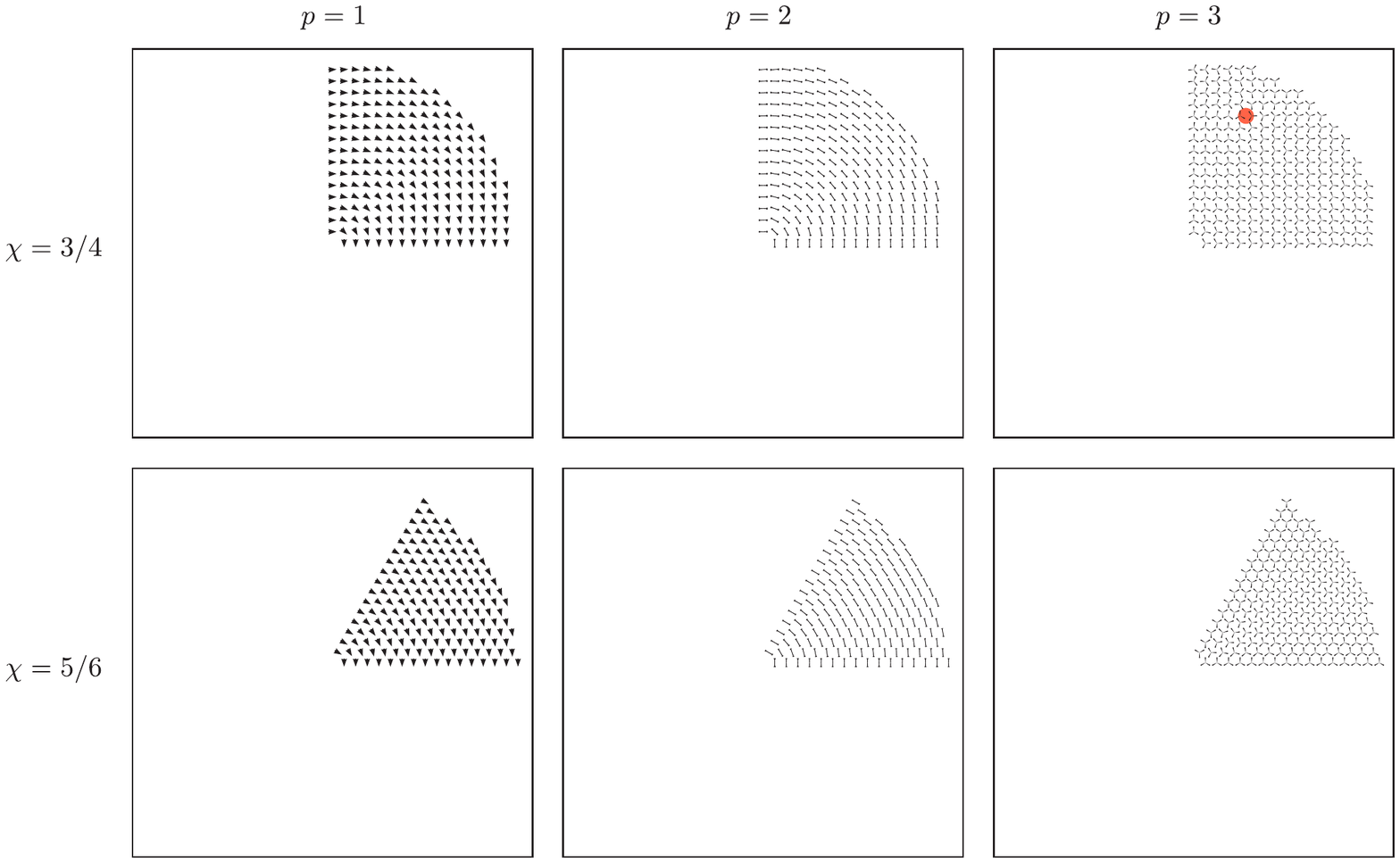}
\end{figure*}

\begin{figure*}
    \centering
    \includegraphics[width=1\textwidth]{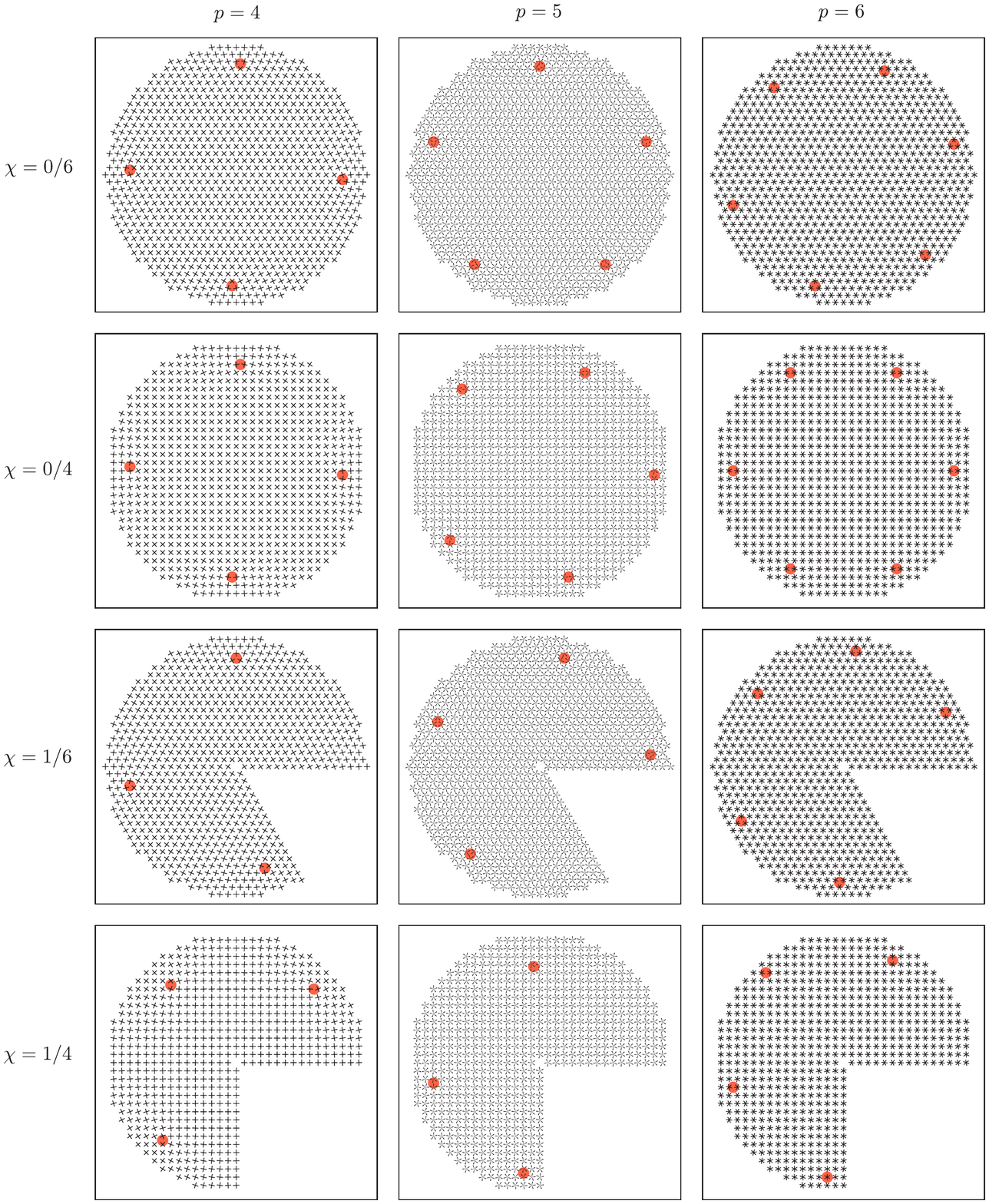}
\end{figure*}

\begin{figure*}
 \includegraphics[width=1\textwidth]{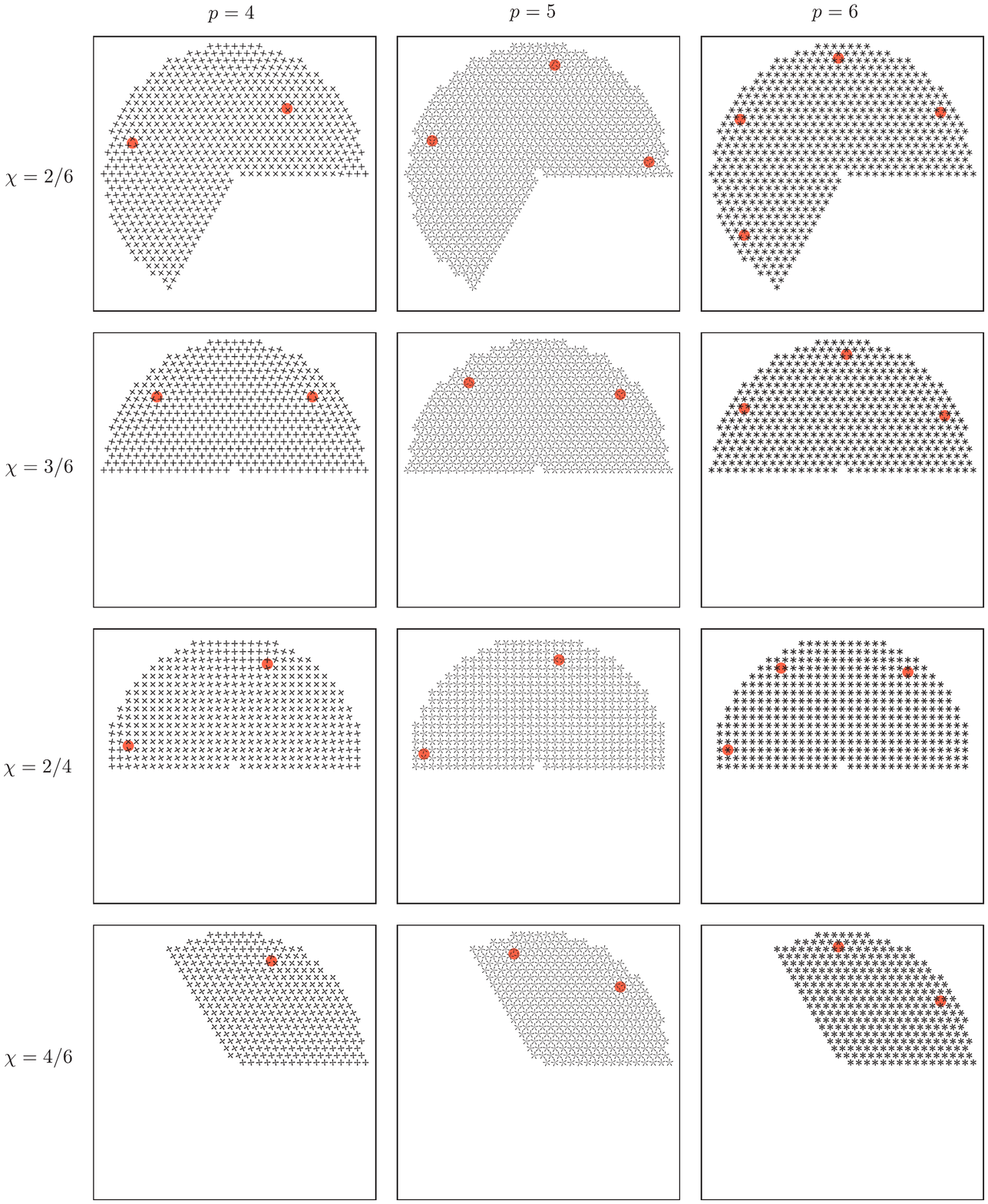}
\end{figure*}

\begin{figure*}
 \includegraphics[width=1\textwidth]{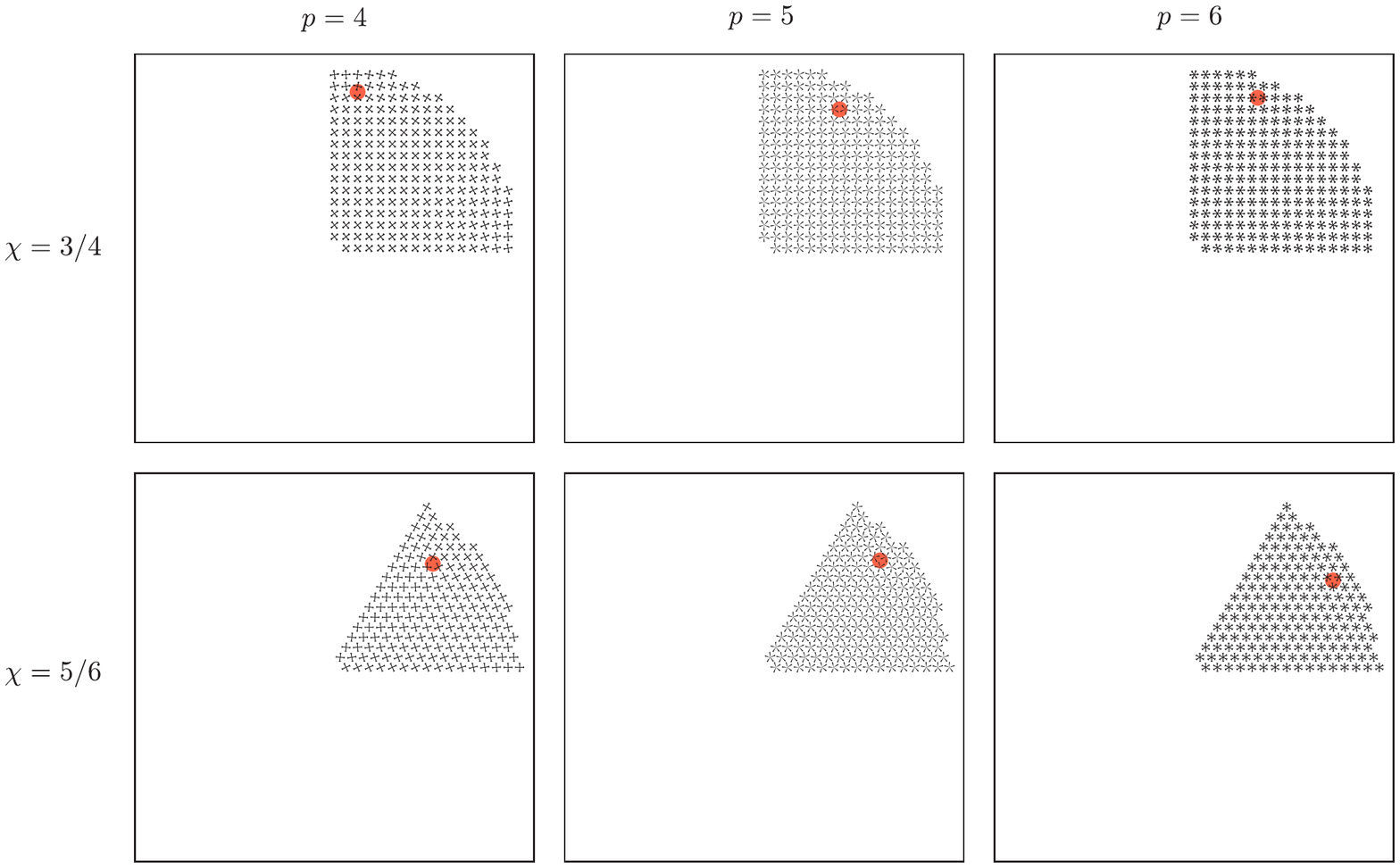}
\end{figure*}
	
	\clearpage
	
	\bibliography{refs}

\begin{thebibliography}{62}%
\makeatletter
\providecommand \@ifxundefined [1]{%
 \@ifx{#1\undefined}
}%
\providecommand \@ifnum [1]{%
 \ifnum #1\expandafter \@firstoftwo
 \else \expandafter \@secondoftwo
 \fi
}%
\providecommand \@ifx [1]{%
 \ifx #1\expandafter \@firstoftwo
 \else \expandafter \@secondoftwo
 \fi
}%
\providecommand \natexlab [1]{#1}%
\providecommand \enquote  [1]{``#1''}%
\providecommand \bibnamefont  [1]{#1}%
\providecommand \bibfnamefont [1]{#1}%
\providecommand \citenamefont [1]{#1}%
\providecommand \href@noop [0]{\@secondoftwo}%
\providecommand \href [0]{\begingroup \@sanitize@url \@href}%
\providecommand \@href[1]{\@@startlink{#1}\@@href}%
\providecommand \@@href[1]{\endgroup#1\@@endlink}%
\providecommand \@sanitize@url [0]{\catcode `\\12\catcode `\$12\catcode
  `\&12\catcode `\#12\catcode `\^12\catcode `\_12\catcode `\%12\relax}%
\providecommand \@@startlink[1]{}%
\providecommand \@@endlink[0]{}%
\providecommand \url  [0]{\begingroup\@sanitize@url \@url }%
\providecommand \@url [1]{\endgroup\@href {#1}{\urlprefix }}%
\providecommand \urlprefix  [0]{URL }%
\providecommand \Eprint [0]{\href }%
\providecommand \doibase [0]{https://doi.org/}%
\providecommand \selectlanguage [0]{\@gobble}%
\providecommand \bibinfo  [0]{\@secondoftwo}%
\providecommand \bibfield  [0]{\@secondoftwo}%
\providecommand \translation [1]{[#1]}%
\providecommand \BibitemOpen [0]{}%
\providecommand \bibitemStop [0]{}%
\providecommand \bibitemNoStop [0]{.\EOS\space}%
\providecommand \EOS [0]{\spacefactor3000\relax}%
\providecommand \BibitemShut  [1]{\csname bibitem#1\endcsname}%
\let\auto@bib@innerbib\@empty
\bibitem [{\citenamefont {Halperin}\ and\ \citenamefont
  {Nelson}(1978)}]{halperin1978theory}%
  \BibitemOpen
  \bibfield  {author} {\bibinfo {author} {\bibfnamefont {B.~I.}\ \bibnamefont
  {Halperin}}\ and\ \bibinfo {author} {\bibfnamefont {D.~R.}\ \bibnamefont
  {Nelson}},\ }\bibfield  {title} {\bibinfo {title} {Theory of two-dimensional
  melting},\ }\href {https://doi.org/10.1103/PhysRevLett.41.121} {\bibfield
  {journal} {\bibinfo  {journal} {Phys. Rev. Lett.}\ }\textbf {\bibinfo
  {volume} {41}},\ \bibinfo {pages} {121} (\bibinfo {year} {1978})}\BibitemShut
  {NoStop}%
\bibitem [{\citenamefont {Nelson}\ and\ \citenamefont
  {Halperin}(1979)}]{nelson1979dislocation}%
  \BibitemOpen
  \bibfield  {author} {\bibinfo {author} {\bibfnamefont {D.~R.}\ \bibnamefont
  {Nelson}}\ and\ \bibinfo {author} {\bibfnamefont {B.~I.}\ \bibnamefont
  {Halperin}},\ }\bibfield  {title} {\bibinfo {title} {Dislocation-mediated
  melting in two dimensions},\ }\href
  {https://doi.org/10.1103/PhysRevB.19.2457} {\bibfield  {journal} {\bibinfo
  {journal} {Phys. Rev. B}\ }\textbf {\bibinfo {volume} {19}},\ \bibinfo
  {pages} {2457} (\bibinfo {year} {1979})}\BibitemShut {NoStop}%
\bibitem [{\citenamefont {Kosterlitz}\ and\ \citenamefont
  {Thouless}(1972)}]{kosterlitz1972long}%
  \BibitemOpen
  \bibfield  {author} {\bibinfo {author} {\bibfnamefont {J.~M.}\ \bibnamefont
  {Kosterlitz}}\ and\ \bibinfo {author} {\bibfnamefont {D.~J.}\ \bibnamefont
  {Thouless}},\ }\bibfield  {title} {\bibinfo {title} {Long range order and
  metastability in two dimensional solids and superfluids. (application of
  dislocation theory)},\ }\href {https://doi.org/10.1088/0022-3719/5/11/002}
  {\bibfield  {journal} {\bibinfo  {journal} {Journal of Physics C: Solid State
  Physics}\ }\textbf {\bibinfo {volume} {5}},\ \bibinfo {pages} {L124}
  (\bibinfo {year} {1972})}\BibitemShut {NoStop}%
\bibitem [{\citenamefont {Kosterlitz}\ and\ \citenamefont
  {Thouless}(1973)}]{kosterlitz1973ordering}%
  \BibitemOpen
  \bibfield  {author} {\bibinfo {author} {\bibfnamefont {J.~M.}\ \bibnamefont
  {Kosterlitz}}\ and\ \bibinfo {author} {\bibfnamefont {D.~J.}\ \bibnamefont
  {Thouless}},\ }\bibfield  {title} {\bibinfo {title} {Ordering, metastability
  and phase transitions in two-dimensional systems},\ }\href
  {https://doi.org/10.1088/0022-3719/6/7/010} {\bibfield  {journal} {\bibinfo
  {journal} {Journal of Physics C: Solid State Physics}\ }\textbf {\bibinfo
  {volume} {6}},\ \bibinfo {pages} {1181} (\bibinfo {year} {1973})}\BibitemShut
  {NoStop}%
\bibitem [{\citenamefont {Young}(1979)}]{young1979melting}%
  \BibitemOpen
  \bibfield  {author} {\bibinfo {author} {\bibfnamefont {A.~P.}\ \bibnamefont
  {Young}},\ }\bibfield  {title} {\bibinfo {title} {Melting and the vector
  coulomb gas in two dimensions},\ }\href
  {https://doi.org/10.1103/PhysRevB.19.1855} {\bibfield  {journal} {\bibinfo
  {journal} {Phys. Rev. B}\ }\textbf {\bibinfo {volume} {19}},\ \bibinfo
  {pages} {1855} (\bibinfo {year} {1979})}\BibitemShut {NoStop}%
\bibitem [{\citenamefont {Li}\ and\ \citenamefont
  {Ciamarra}(2018)}]{li2018role}%
  \BibitemOpen
  \bibfield  {author} {\bibinfo {author} {\bibfnamefont {Y.-W.}\ \bibnamefont
  {Li}}\ and\ \bibinfo {author} {\bibfnamefont {M.~P.}\ \bibnamefont
  {Ciamarra}},\ }\bibfield  {title} {\bibinfo {title} {Role of cell
  deformability in the two-dimensional melting of biological tissues},\ }\href
  {https://doi.org/10.1103/PhysRevMaterials.2.045602} {\bibfield  {journal}
  {\bibinfo  {journal} {Phys. Rev. Materials}\ }\textbf {\bibinfo {volume}
  {2}},\ \bibinfo {pages} {045602} (\bibinfo {year} {2018})}\BibitemShut
  {NoStop}%
\bibitem [{\citenamefont {Korolev}\ and\ \citenamefont
  {Nelson}(2008)}]{korolev2008defect}%
  \BibitemOpen
  \bibfield  {author} {\bibinfo {author} {\bibfnamefont {K.~S.}\ \bibnamefont
  {Korolev}}\ and\ \bibinfo {author} {\bibfnamefont {D.~R.}\ \bibnamefont
  {Nelson}},\ }\bibfield  {title} {\bibinfo {title} {Defect-mediated
  emulsification in two dimensions},\ }\href
  {https://doi.org/10.1103/PhysRevE.77.051702} {\bibfield  {journal} {\bibinfo
  {journal} {Phys. Rev. E}\ }\textbf {\bibinfo {volume} {77}},\ \bibinfo
  {pages} {051702} (\bibinfo {year} {2008})}\BibitemShut {NoStop}%
\bibitem [{\citenamefont {Dimova}\ \emph {et~al.}(2000)\citenamefont {Dimova},
  \citenamefont {Pouligny},\ and\ \citenamefont
  {Dietrich}}]{dimova2000pretransitional}%
  \BibitemOpen
  \bibfield  {author} {\bibinfo {author} {\bibfnamefont {R.}~\bibnamefont
  {Dimova}}, \bibinfo {author} {\bibfnamefont {B.}~\bibnamefont {Pouligny}},\
  and\ \bibinfo {author} {\bibfnamefont {C.}~\bibnamefont {Dietrich}},\
  }\bibfield  {title} {\bibinfo {title} {Pretransitional effects in
  dimyristoylphosphatidylcholine vesicle membranes: optical dynamometry
  study},\ }\href@noop {} {\bibfield  {journal} {\bibinfo  {journal}
  {Biophysical Journal}\ }\textbf {\bibinfo {volume} {79}},\ \bibinfo {pages}
  {340} (\bibinfo {year} {2000})}\BibitemShut {NoStop}%
\bibitem [{\citenamefont {de~Gennes}\ and\ \citenamefont
  {Prost}(1993)}]{gennes1993the}%
  \BibitemOpen
  \bibfield  {author} {\bibinfo {author} {\bibfnamefont {P.~G.}\ \bibnamefont
  {de~Gennes}}\ and\ \bibinfo {author} {\bibfnamefont {J.}~\bibnamefont
  {Prost}},\ }\href@noop {} {\emph {\bibinfo {title} {The physics of liquid
  crystals}}}\ (\bibinfo  {publisher} {Clarendon Press, Oxford},\ \bibinfo
  {year} {1993})\BibitemShut {NoStop}%
\bibitem [{\citenamefont {Thorneywork}\ \emph {et~al.}(2017)\citenamefont
  {Thorneywork}, \citenamefont {Abbott}, \citenamefont {Aarts},\ and\
  \citenamefont {Dullens}}]{thorneywork2017two}%
  \BibitemOpen
  \bibfield  {author} {\bibinfo {author} {\bibfnamefont {A.~L.}\ \bibnamefont
  {Thorneywork}}, \bibinfo {author} {\bibfnamefont {J.~L.}\ \bibnamefont
  {Abbott}}, \bibinfo {author} {\bibfnamefont {D.~G. A.~L.}\ \bibnamefont
  {Aarts}},\ and\ \bibinfo {author} {\bibfnamefont {R.~P.~A.}\ \bibnamefont
  {Dullens}},\ }\bibfield  {title} {\bibinfo {title} {Two-dimensional melting
  of colloidal hard spheres},\ }\href
  {https://doi.org/10.1103/PhysRevLett.118.158001} {\bibfield  {journal}
  {\bibinfo  {journal} {Phys. Rev. Lett.}\ }\textbf {\bibinfo {volume} {118}},\
  \bibinfo {pages} {158001} (\bibinfo {year} {2017})}\BibitemShut {NoStop}%
\bibitem [{\citenamefont {Zhao}\ \emph {et~al.}(2012)\citenamefont {Zhao},
  \citenamefont {Bruinsma},\ and\ \citenamefont {Mason}}]{zhao2012local}%
  \BibitemOpen
  \bibfield  {author} {\bibinfo {author} {\bibfnamefont {K.}~\bibnamefont
  {Zhao}}, \bibinfo {author} {\bibfnamefont {R.}~\bibnamefont {Bruinsma}},\
  and\ \bibinfo {author} {\bibfnamefont {T.~G.}\ \bibnamefont {Mason}},\
  }\bibfield  {title} {\bibinfo {title} {Local chiral symmetry breaking in
  triatic liquid crystals},\ }\href {https://doi.org/10.1038/ncomms1803}
  {\bibfield  {journal} {\bibinfo  {journal} {Nature Communications}\ }\textbf
  {\bibinfo {volume} {3}},\ \bibinfo {pages} {801} (\bibinfo {year}
  {2012})}\BibitemShut {NoStop}%
\bibitem [{\citenamefont {Löffler}(2018)}]{loffler2018phase}%
  \BibitemOpen
  \bibfield  {author} {\bibinfo {author} {\bibfnamefont {R.~C.}\ \bibnamefont
  {Löffler}},\ }\emph {\bibinfo {title} {Phase behavior of 2D monolayers of
  cubic colloids}},\ \href@noop {} {Master's thesis},\ \bibinfo  {school}
  {Universität Konstanz}, \bibinfo {address} {Konstanz} (\bibinfo {year}
  {2018})\BibitemShut {NoStop}%
\bibitem [{\citenamefont {Mietke}\ and\ \citenamefont
  {Dunkel}(2022)}]{mietke2022anyonic}%
  \BibitemOpen
  \bibfield  {author} {\bibinfo {author} {\bibfnamefont {A.}~\bibnamefont
  {Mietke}}\ and\ \bibinfo {author} {\bibfnamefont {J.}~\bibnamefont
  {Dunkel}},\ }\bibfield  {title} {\bibinfo {title} {Anyonic defect braiding
  and spontaneous chiral symmetry breaking in dihedral liquid crystals},\
  }\href {https://doi.org/10.1103/PhysRevX.12.011027} {\bibfield  {journal}
  {\bibinfo  {journal} {Phys. Rev. X}\ }\textbf {\bibinfo {volume} {12}},\
  \bibinfo {pages} {011027} (\bibinfo {year} {2022})}\BibitemShut {NoStop}%
\bibitem [{\citenamefont {Marchetti}\ \emph {et~al.}(2013)\citenamefont
  {Marchetti}, \citenamefont {Joanny}, \citenamefont {Ramaswamy}, \citenamefont
  {Liverpool}, \citenamefont {Prost}, \citenamefont {Rao},\ and\ \citenamefont
  {Simha}}]{marchetti2013hydrodynamics}%
  \BibitemOpen
  \bibfield  {author} {\bibinfo {author} {\bibfnamefont {M.~C.}\ \bibnamefont
  {Marchetti}}, \bibinfo {author} {\bibfnamefont {J.~F.}\ \bibnamefont
  {Joanny}}, \bibinfo {author} {\bibfnamefont {S.}~\bibnamefont {Ramaswamy}},
  \bibinfo {author} {\bibfnamefont {T.~B.}\ \bibnamefont {Liverpool}}, \bibinfo
  {author} {\bibfnamefont {J.}~\bibnamefont {Prost}}, \bibinfo {author}
  {\bibfnamefont {M.}~\bibnamefont {Rao}},\ and\ \bibinfo {author}
  {\bibfnamefont {R.~A.}\ \bibnamefont {Simha}},\ }\bibfield  {title} {\bibinfo
  {title} {Hydrodynamics of soft active matter},\ }\href
  {https://doi.org/10.1103/RevModPhys.85.1143} {\bibfield  {journal} {\bibinfo
  {journal} {Rev. Mod. Phys.}\ }\textbf {\bibinfo {volume} {85}},\ \bibinfo
  {pages} {1143} (\bibinfo {year} {2013})}\BibitemShut {NoStop}%
\bibitem [{\citenamefont {Vicsek}\ \emph {et~al.}(1995)\citenamefont {Vicsek},
  \citenamefont {Czir\'ok}, \citenamefont {Ben-Jacob}, \citenamefont {Cohen},\
  and\ \citenamefont {Shochet}}]{vicsek1995novel}%
  \BibitemOpen
  \bibfield  {author} {\bibinfo {author} {\bibfnamefont {T.}~\bibnamefont
  {Vicsek}}, \bibinfo {author} {\bibfnamefont {A.}~\bibnamefont {Czir\'ok}},
  \bibinfo {author} {\bibfnamefont {E.}~\bibnamefont {Ben-Jacob}}, \bibinfo
  {author} {\bibfnamefont {I.}~\bibnamefont {Cohen}},\ and\ \bibinfo {author}
  {\bibfnamefont {O.}~\bibnamefont {Shochet}},\ }\bibfield  {title} {\bibinfo
  {title} {Novel type of phase transition in a system of self-driven
  particles},\ }\href {https://doi.org/10.1103/PhysRevLett.75.1226} {\bibfield
  {journal} {\bibinfo  {journal} {Phys. Rev. Lett.}\ }\textbf {\bibinfo
  {volume} {75}},\ \bibinfo {pages} {1226} (\bibinfo {year}
  {1995})}\BibitemShut {NoStop}%
\bibitem [{\citenamefont {Toner}\ and\ \citenamefont
  {Tu}(1995)}]{toner1995long}%
  \BibitemOpen
  \bibfield  {author} {\bibinfo {author} {\bibfnamefont {J.}~\bibnamefont
  {Toner}}\ and\ \bibinfo {author} {\bibfnamefont {Y.}~\bibnamefont {Tu}},\
  }\bibfield  {title} {\bibinfo {title} {Long-range order in a two-dimensional
  dynamical $\mathrm{XY}$ model: How birds fly together},\ }\href
  {https://doi.org/10.1103/PhysRevLett.75.4326} {\bibfield  {journal} {\bibinfo
   {journal} {Phys. Rev. Lett.}\ }\textbf {\bibinfo {volume} {75}},\ \bibinfo
  {pages} {4326} (\bibinfo {year} {1995})}\BibitemShut {NoStop}%
\bibitem [{\citenamefont {Toner}\ and\ \citenamefont
  {Tu}(1998)}]{toner1998flocks}%
  \BibitemOpen
  \bibfield  {author} {\bibinfo {author} {\bibfnamefont {J.}~\bibnamefont
  {Toner}}\ and\ \bibinfo {author} {\bibfnamefont {Y.}~\bibnamefont {Tu}},\
  }\bibfield  {title} {\bibinfo {title} {Flocks, herds, and schools: A
  quantitative theory of flocking},\ }\href
  {https://doi.org/10.1103/PhysRevE.58.4828} {\bibfield  {journal} {\bibinfo
  {journal} {Phys. Rev. E}\ }\textbf {\bibinfo {volume} {58}},\ \bibinfo
  {pages} {4828} (\bibinfo {year} {1998})}\BibitemShut {NoStop}%
\bibitem [{\citenamefont {Toner}\ \emph {et~al.}(2005)\citenamefont {Toner},
  \citenamefont {Tu},\ and\ \citenamefont
  {Ramaswamy}}]{toner2005hydrodynamics}%
  \BibitemOpen
  \bibfield  {author} {\bibinfo {author} {\bibfnamefont {J.}~\bibnamefont
  {Toner}}, \bibinfo {author} {\bibfnamefont {Y.}~\bibnamefont {Tu}},\ and\
  \bibinfo {author} {\bibfnamefont {S.}~\bibnamefont {Ramaswamy}},\ }\bibfield
  {title} {\bibinfo {title} {Hydrodynamics and phases of flocks},\ }\href
  {https://doi.org/https://doi.org/10.1016/j.aop.2005.04.011} {\bibfield
  {journal} {\bibinfo  {journal} {Ann. Phys.}\ }\textbf {\bibinfo {volume}
  {318}},\ \bibinfo {pages} {170 } (\bibinfo {year} {2005})}\BibitemShut
  {NoStop}%
\bibitem [{\citenamefont {Wensink}\ \emph {et~al.}(2012)\citenamefont
  {Wensink}, \citenamefont {Dunkel}, \citenamefont {Heidenreich}, \citenamefont
  {Drescher}, \citenamefont {Goldstein}, \citenamefont {L{\"o}wen},\ and\
  \citenamefont {Yeomans}}]{wensink2012meso}%
  \BibitemOpen
  \bibfield  {author} {\bibinfo {author} {\bibfnamefont {H.~H.}\ \bibnamefont
  {Wensink}}, \bibinfo {author} {\bibfnamefont {J.}~\bibnamefont {Dunkel}},
  \bibinfo {author} {\bibfnamefont {S.}~\bibnamefont {Heidenreich}}, \bibinfo
  {author} {\bibfnamefont {K.}~\bibnamefont {Drescher}}, \bibinfo {author}
  {\bibfnamefont {R.~E.}\ \bibnamefont {Goldstein}}, \bibinfo {author}
  {\bibfnamefont {H.}~\bibnamefont {L{\"o}wen}},\ and\ \bibinfo {author}
  {\bibfnamefont {J.~M.}\ \bibnamefont {Yeomans}},\ }\bibfield  {title}
  {\bibinfo {title} {Meso-scale turbulence in living fluids},\ }\href@noop {}
  {\bibfield  {journal} {\bibinfo  {journal} {PNAS}\ }\textbf {\bibinfo
  {volume} {109}},\ \bibinfo {pages} {14308} (\bibinfo {year}
  {2012})}\BibitemShut {NoStop}%
\bibitem [{\citenamefont {Bricard}\ \emph {et~al.}(2013)\citenamefont
  {Bricard}, \citenamefont {Caussin}, \citenamefont {Desreumaux}, \citenamefont
  {Dauchot},\ and\ \citenamefont {Bartolo}}]{bricard2013emergence}%
  \BibitemOpen
  \bibfield  {author} {\bibinfo {author} {\bibfnamefont {A.}~\bibnamefont
  {Bricard}}, \bibinfo {author} {\bibfnamefont {J.-B.}\ \bibnamefont
  {Caussin}}, \bibinfo {author} {\bibfnamefont {N.}~\bibnamefont {Desreumaux}},
  \bibinfo {author} {\bibfnamefont {O.}~\bibnamefont {Dauchot}},\ and\ \bibinfo
  {author} {\bibfnamefont {D.}~\bibnamefont {Bartolo}},\ }\bibfield  {title}
  {\bibinfo {title} {Emergence of macroscopic directed motion in populations of
  motile colloids},\ }\href {https://doi.org/10.1038/nature12673} {\bibfield
  {journal} {\bibinfo  {journal} {Nature}\ }\textbf {\bibinfo {volume} {503}},\
  \bibinfo {pages} {95–98} (\bibinfo {year} {2013})}\BibitemShut {NoStop}%
\bibitem [{\citenamefont {Aditi~Simha}\ and\ \citenamefont
  {Ramaswamy}(2002)}]{simha2002hydrodynamic}%
  \BibitemOpen
  \bibfield  {author} {\bibinfo {author} {\bibfnamefont {R.}~\bibnamefont
  {Aditi~Simha}}\ and\ \bibinfo {author} {\bibfnamefont {S.}~\bibnamefont
  {Ramaswamy}},\ }\bibfield  {title} {\bibinfo {title} {Hydrodynamic
  fluctuations and instabilities in ordered suspensions of self-propelled
  particles},\ }\href {https://doi.org/10.1103/PhysRevLett.89.058101}
  {\bibfield  {journal} {\bibinfo  {journal} {Phys. Rev. Lett.}\ }\textbf
  {\bibinfo {volume} {89}},\ \bibinfo {pages} {058101} (\bibinfo {year}
  {2002})}\BibitemShut {NoStop}%
\bibitem [{\citenamefont {Doostmohammadi}\ \emph {et~al.}(2018)\citenamefont
  {Doostmohammadi}, \citenamefont {Ign{\'e}s-Mullol}, \citenamefont {Yeomans},\
  and\ \citenamefont {Sagu{\'e}s}}]{doostmohammadi2018active}%
  \BibitemOpen
  \bibfield  {author} {\bibinfo {author} {\bibfnamefont {A.}~\bibnamefont
  {Doostmohammadi}}, \bibinfo {author} {\bibfnamefont {J.}~\bibnamefont
  {Ign{\'e}s-Mullol}}, \bibinfo {author} {\bibfnamefont {J.~M.}\ \bibnamefont
  {Yeomans}},\ and\ \bibinfo {author} {\bibfnamefont {F.}~\bibnamefont
  {Sagu{\'e}s}},\ }\bibfield  {title} {\bibinfo {title} {Active nematics},\
  }\href {https://doi.org/10.1038/s41467-018-05666-8} {\bibfield  {journal}
  {\bibinfo  {journal} {Nat. Comm.}\ }\textbf {\bibinfo {volume} {9}},\
  \bibinfo {pages} {3246} (\bibinfo {year} {2018})}\BibitemShut {NoStop}%
\bibitem [{\citenamefont {Duclos}\ \emph {et~al.}(2016)\citenamefont {Duclos},
  \citenamefont {Erlenk{\"a}mper}, \citenamefont {Joanny},\ and\ \citenamefont
  {Silberzan}}]{duclos2017topological}%
  \BibitemOpen
  \bibfield  {author} {\bibinfo {author} {\bibfnamefont {G.}~\bibnamefont
  {Duclos}}, \bibinfo {author} {\bibfnamefont {C.}~\bibnamefont
  {Erlenk{\"a}mper}}, \bibinfo {author} {\bibfnamefont {J.-F.}\ \bibnamefont
  {Joanny}},\ and\ \bibinfo {author} {\bibfnamefont {P.}~\bibnamefont
  {Silberzan}},\ }\bibfield  {title} {\bibinfo {title} {Topological defects in
  confined populations of spindle-shaped cells},\ }\href
  {https://doi.org/10.1038/nphys3876} {\bibfield  {journal} {\bibinfo
  {journal} {Nat. Phys.}\ }\textbf {\bibinfo {volume} {13}},\ \bibinfo {pages}
  {58} (\bibinfo {year} {2016})}\BibitemShut {NoStop}%
\bibitem [{\citenamefont {Kawaguchi}\ \emph {et~al.}(2017)\citenamefont
  {Kawaguchi}, \citenamefont {Kageyama},\ and\ \citenamefont
  {Sano}}]{kawaguchi2017topological}%
  \BibitemOpen
  \bibfield  {author} {\bibinfo {author} {\bibfnamefont {K.}~\bibnamefont
  {Kawaguchi}}, \bibinfo {author} {\bibfnamefont {R.}~\bibnamefont
  {Kageyama}},\ and\ \bibinfo {author} {\bibfnamefont {M.}~\bibnamefont
  {Sano}},\ }\bibfield  {title} {\bibinfo {title} {Topological defects control
  collective dynamics in neural progenitor cell cultures},\ }\href
  {https://doi.org/10.1038/nature22321} {\bibfield  {journal} {\bibinfo
  {journal} {Nature}\ }\textbf {\bibinfo {volume} {545}},\ \bibinfo {pages}
  {327–331} (\bibinfo {year} {2017})}\BibitemShut {NoStop}%
\bibitem [{\citenamefont {Saw}\ \emph {et~al.}(2017)\citenamefont {Saw},
  \citenamefont {Doostmohammadi}, \citenamefont {Nier}, \citenamefont
  {Kocgozlu}, \citenamefont {Thampi}, \citenamefont {Toyama}, \citenamefont
  {Marcq}, \citenamefont {Lim}, \citenamefont {Yeomans},\ and\ \citenamefont
  {Ladoux}}]{saw2017topological}%
  \BibitemOpen
  \bibfield  {author} {\bibinfo {author} {\bibfnamefont {T.~B.}\ \bibnamefont
  {Saw}}, \bibinfo {author} {\bibfnamefont {A.}~\bibnamefont {Doostmohammadi}},
  \bibinfo {author} {\bibfnamefont {V.}~\bibnamefont {Nier}}, \bibinfo {author}
  {\bibfnamefont {L.}~\bibnamefont {Kocgozlu}}, \bibinfo {author}
  {\bibfnamefont {S.}~\bibnamefont {Thampi}}, \bibinfo {author} {\bibfnamefont
  {Y.}~\bibnamefont {Toyama}}, \bibinfo {author} {\bibfnamefont
  {P.}~\bibnamefont {Marcq}}, \bibinfo {author} {\bibfnamefont {C.~T.}\
  \bibnamefont {Lim}}, \bibinfo {author} {\bibfnamefont {J.~M.}\ \bibnamefont
  {Yeomans}},\ and\ \bibinfo {author} {\bibfnamefont {B.}~\bibnamefont
  {Ladoux}},\ }\bibfield  {title} {\bibinfo {title} {Topological defects in
  epithelia govern cell death and extrusion},\ }\href
  {https://doi.org/10.1038/nature21718} {\bibfield  {journal} {\bibinfo
  {journal} {Nature}\ }\textbf {\bibinfo {volume} {544}},\ \bibinfo {pages}
  {212–216} (\bibinfo {year} {2017})}\BibitemShut {NoStop}%
\bibitem [{\citenamefont {Blanch-Mercader}\ \emph {et~al.}(2018)\citenamefont
  {Blanch-Mercader}, \citenamefont {Yashunsky}, \citenamefont {Garcia},
  \citenamefont {Duclos}, \citenamefont {Giomi},\ and\ \citenamefont
  {Silberzan}}]{blanch2018turbulent}%
  \BibitemOpen
  \bibfield  {author} {\bibinfo {author} {\bibfnamefont {C.}~\bibnamefont
  {Blanch-Mercader}}, \bibinfo {author} {\bibfnamefont {V.}~\bibnamefont
  {Yashunsky}}, \bibinfo {author} {\bibfnamefont {S.}~\bibnamefont {Garcia}},
  \bibinfo {author} {\bibfnamefont {G.}~\bibnamefont {Duclos}}, \bibinfo
  {author} {\bibfnamefont {L.}~\bibnamefont {Giomi}},\ and\ \bibinfo {author}
  {\bibfnamefont {P.}~\bibnamefont {Silberzan}},\ }\bibfield  {title} {\bibinfo
  {title} {Turbulent dynamics of epithelial cell cultures},\ }\href
  {https://doi.org/10.1103/PhysRevLett.120.208101} {\bibfield  {journal}
  {\bibinfo  {journal} {Phys. Rev. Lett.}\ }\textbf {\bibinfo {volume} {120}},\
  \bibinfo {pages} {208101} (\bibinfo {year} {2018})}\BibitemShut {NoStop}%
\bibitem [{\citenamefont {Sanchez}\ \emph {et~al.}(2012)\citenamefont
  {Sanchez}, \citenamefont {Chen}, \citenamefont {DeCamp}, \citenamefont
  {Heymann},\ and\ \citenamefont {Dogic}}]{sanchez2012spontaneous}%
  \BibitemOpen
  \bibfield  {author} {\bibinfo {author} {\bibfnamefont {T.}~\bibnamefont
  {Sanchez}}, \bibinfo {author} {\bibfnamefont {D.~T.~N.}\ \bibnamefont
  {Chen}}, \bibinfo {author} {\bibfnamefont {S.~J.}\ \bibnamefont {DeCamp}},
  \bibinfo {author} {\bibfnamefont {M.}~\bibnamefont {Heymann}},\ and\ \bibinfo
  {author} {\bibfnamefont {Z.}~\bibnamefont {Dogic}},\ }\bibfield  {title}
  {\bibinfo {title} {Spontaneous motion in hierarchically assembled active
  matter},\ }\href {https://doi.org/10.1038/nature11591} {\bibfield  {journal}
  {\bibinfo  {journal} {Nature}\ }\textbf {\bibinfo {volume} {491}},\ \bibinfo
  {pages} {431} (\bibinfo {year} {2012})}\BibitemShut {NoStop}%
\bibitem [{\citenamefont {Keber}\ \emph {et~al.}(2014)\citenamefont {Keber},
  \citenamefont {Loiseau}, \citenamefont {Sanchez}, \citenamefont {DeCamp},
  \citenamefont {Giomi}, \citenamefont {Bowick}, \citenamefont {Marchetti},
  \citenamefont {Dogic},\ and\ \citenamefont {Bausch}}]{keber2014topology}%
  \BibitemOpen
  \bibfield  {author} {\bibinfo {author} {\bibfnamefont {F.~C.}\ \bibnamefont
  {Keber}}, \bibinfo {author} {\bibfnamefont {E.}~\bibnamefont {Loiseau}},
  \bibinfo {author} {\bibfnamefont {T.}~\bibnamefont {Sanchez}}, \bibinfo
  {author} {\bibfnamefont {S.~J.}\ \bibnamefont {DeCamp}}, \bibinfo {author}
  {\bibfnamefont {L.}~\bibnamefont {Giomi}}, \bibinfo {author} {\bibfnamefont
  {M.~J.}\ \bibnamefont {Bowick}}, \bibinfo {author} {\bibfnamefont {M.~C.}\
  \bibnamefont {Marchetti}}, \bibinfo {author} {\bibfnamefont {Z.}~\bibnamefont
  {Dogic}},\ and\ \bibinfo {author} {\bibfnamefont {A.~R.}\ \bibnamefont
  {Bausch}},\ }\bibfield  {title} {\bibinfo {title} {Topology and dynamics of
  active nematic vesicles},\ }\href {https://doi.org/10.1126/science.1254784}
  {\bibfield  {journal} {\bibinfo  {journal} {Science}\ }\textbf {\bibinfo
  {volume} {345}},\ \bibinfo {pages} {1135} (\bibinfo {year}
  {2014})}\BibitemShut {NoStop}%
\bibitem [{\citenamefont {Kumar}\ \emph {et~al.}(2018)\citenamefont {Kumar},
  \citenamefont {Zhang}, \citenamefont {de~Pablo},\ and\ \citenamefont
  {Gardel}}]{kumar2018tunable}%
  \BibitemOpen
  \bibfield  {author} {\bibinfo {author} {\bibfnamefont {N.}~\bibnamefont
  {Kumar}}, \bibinfo {author} {\bibfnamefont {R.}~\bibnamefont {Zhang}},
  \bibinfo {author} {\bibfnamefont {J.~J.}\ \bibnamefont {de~Pablo}},\ and\
  \bibinfo {author} {\bibfnamefont {M.~L.}\ \bibnamefont {Gardel}},\ }\bibfield
   {title} {\bibinfo {title} {Tunable structure and dynamics of active liquid
  crystals},\ }\href {https://doi.org/10.1126/sciadv.aat7779} {\bibfield
  {journal} {\bibinfo  {journal} {Sci. Adv.}\ }\textbf {\bibinfo {volume}
  {4}},\ \bibinfo {eid} {eaat7779} (\bibinfo {year} {2018})}\BibitemShut
  {NoStop}%
\bibitem [{\citenamefont {Doostmohammadi}\ \emph {et~al.}(2016)\citenamefont
  {Doostmohammadi}, \citenamefont {Thampi},\ and\ \citenamefont
  {Yeomans}}]{doostmohammadi2016defect}%
  \BibitemOpen
  \bibfield  {author} {\bibinfo {author} {\bibfnamefont {A.}~\bibnamefont
  {Doostmohammadi}}, \bibinfo {author} {\bibfnamefont {S.~P.}\ \bibnamefont
  {Thampi}},\ and\ \bibinfo {author} {\bibfnamefont {J.~M.}\ \bibnamefont
  {Yeomans}},\ }\bibfield  {title} {\bibinfo {title} {Defect-mediated
  morphologies in growing cell colonies},\ }\href@noop {} {\bibfield  {journal}
  {\bibinfo  {journal} {Phys. Rev. Lett.}\ }\textbf {\bibinfo {volume} {117}},\
  \bibinfo {pages} {048102} (\bibinfo {year} {2016})}\BibitemShut {NoStop}%
\bibitem [{\citenamefont {Nishiguchi}\ \emph {et~al.}(2017)\citenamefont
  {Nishiguchi}, \citenamefont {Nagai}, \citenamefont {Chat\'e},\ and\
  \citenamefont {Sano}}]{nishiguchi2017long-range}%
  \BibitemOpen
  \bibfield  {author} {\bibinfo {author} {\bibfnamefont {D.}~\bibnamefont
  {Nishiguchi}}, \bibinfo {author} {\bibfnamefont {K.~H.}\ \bibnamefont
  {Nagai}}, \bibinfo {author} {\bibfnamefont {H.}~\bibnamefont {Chat\'e}},\
  and\ \bibinfo {author} {\bibfnamefont {M.}~\bibnamefont {Sano}},\ }\bibfield
  {title} {\bibinfo {title} {Long-range nematic order and anomalous
  fluctuations in suspensions of swimming filamentous bacteria},\ }\href
  {https://doi.org/10.1103/PhysRevE.95.020601} {\bibfield  {journal} {\bibinfo
  {journal} {Phys. Rev. E}\ }\textbf {\bibinfo {volume} {95}},\ \bibinfo
  {pages} {020601} (\bibinfo {year} {2017})}\BibitemShut {NoStop}%
\bibitem [{\citenamefont {Copenhagen}\ \emph {et~al.}(2021)\citenamefont
  {Copenhagen}, \citenamefont {Alert}, \citenamefont {Wingreen},\ and\
  \citenamefont {Shaevitz}}]{copenhagen2020topological}%
  \BibitemOpen
  \bibfield  {author} {\bibinfo {author} {\bibfnamefont {K.}~\bibnamefont
  {Copenhagen}}, \bibinfo {author} {\bibfnamefont {R.}~\bibnamefont {Alert}},
  \bibinfo {author} {\bibfnamefont {N.~S.}\ \bibnamefont {Wingreen}},\ and\
  \bibinfo {author} {\bibfnamefont {J.~W.}\ \bibnamefont {Shaevitz}},\
  }\bibfield  {title} {\bibinfo {title} {Topological defects promote layer
  formation in {Myxococcus} xanthus colonies},\ }\href
  {https://doi.org/10.1038/s41567-020-01056-4} {\bibfield  {journal} {\bibinfo
  {journal} {Nature Physics}\ }\textbf {\bibinfo {volume} {17}},\ \bibinfo
  {pages} {211} (\bibinfo {year} {2021})}\BibitemShut {NoStop}%
\bibitem [{\citenamefont {Narayan}\ \emph {et~al.}(2007)\citenamefont
  {Narayan}, \citenamefont {Ramaswamy},\ and\ \citenamefont
  {Menon}}]{narayan2007long}%
  \BibitemOpen
  \bibfield  {author} {\bibinfo {author} {\bibfnamefont {V.}~\bibnamefont
  {Narayan}}, \bibinfo {author} {\bibfnamefont {S.}~\bibnamefont {Ramaswamy}},\
  and\ \bibinfo {author} {\bibfnamefont {N.}~\bibnamefont {Menon}},\ }\bibfield
   {title} {\bibinfo {title} {Long-lived giant number fluctuations in a
  swarming granular nematic},\ }\href {https://doi.org/10.1126/science.1140414}
  {\bibfield  {journal} {\bibinfo  {journal} {Science}\ }\textbf {\bibinfo
  {volume} {317}},\ \bibinfo {pages} {105} (\bibinfo {year}
  {2007})}\BibitemShut {NoStop}%
\bibitem [{\citenamefont {Maroudas-Sacks}\ \emph {et~al.}(2020)\citenamefont
  {Maroudas-Sacks}, \citenamefont {Garion}, \citenamefont {Shani-Zerbib},
  \citenamefont {Livshits}, \citenamefont {Braun},\ and\ \citenamefont
  {Keren}}]{maroudas2020topological}%
  \BibitemOpen
  \bibfield  {author} {\bibinfo {author} {\bibfnamefont {Y.}~\bibnamefont
  {Maroudas-Sacks}}, \bibinfo {author} {\bibfnamefont {L.}~\bibnamefont
  {Garion}}, \bibinfo {author} {\bibfnamefont {L.}~\bibnamefont
  {Shani-Zerbib}}, \bibinfo {author} {\bibfnamefont {A.}~\bibnamefont
  {Livshits}}, \bibinfo {author} {\bibfnamefont {E.}~\bibnamefont {Braun}},\
  and\ \bibinfo {author} {\bibfnamefont {K.}~\bibnamefont {Keren}},\ }\bibfield
   {title} {\bibinfo {title} {Topological defects in the nematic order of actin
  fibres as organization centres of hydra morphogenesis},\ }\bibfield
  {journal} {\bibinfo  {journal} {Nature Physics}\ }\href
  {https://doi.org/10.1038/s41567-020-01083-1} {10.1038/s41567-020-01083-1}
  (\bibinfo {year} {2020})\BibitemShut {NoStop}%
\bibitem [{\citenamefont {Cislo}\ \emph {et~al.}(2021)\citenamefont {Cislo},
  \citenamefont {Qin}, \citenamefont {Yang}, \citenamefont {Bowick},\ and\
  \citenamefont {Streichan}}]{cislo2021active}%
  \BibitemOpen
  \bibfield  {author} {\bibinfo {author} {\bibfnamefont {D.}~\bibnamefont
  {Cislo}}, \bibinfo {author} {\bibfnamefont {H.}~\bibnamefont {Qin}}, \bibinfo
  {author} {\bibfnamefont {F.}~\bibnamefont {Yang}}, \bibinfo {author}
  {\bibfnamefont {M.~J.}\ \bibnamefont {Bowick}},\ and\ \bibinfo {author}
  {\bibfnamefont {S.~J.}\ \bibnamefont {Streichan}},\ }\bibfield  {title}
  {\bibinfo {title} {Active cell divisions generate exotic fourfold
  orientationally ordered phase in living tissue},\ }\bibfield  {journal}
  {\bibinfo  {journal} {bioRxiv}\ }\href
  {https://doi.org/10.1101/2021.07.28.453899} {10.1101/2021.07.28.453899}
  (\bibinfo {year} {2021}),\ \Eprint
  {https://arxiv.org/abs/https://www.biorxiv.org/content/early/2021/07/28/2021.07.28.453899.full.pdf}
  {https://www.biorxiv.org/content/early/2021/07/28/2021.07.28.453899.full.pdf}
  \BibitemShut {NoStop}%
\bibitem [{\citenamefont {{Nelson, D.R.}}\ and\ \citenamefont {{Peliti,
  L.}}(1987)}]{nelson1987fluctuations}%
  \BibitemOpen
  \bibfield  {author} {\bibinfo {author} {\bibnamefont {{Nelson, D.R.}}}\ and\
  \bibinfo {author} {\bibnamefont {{Peliti, L.}}},\ }\bibfield  {title}
  {\bibinfo {title} {Fluctuations in membranes with crystalline and hexatic
  order},\ }\href {https://doi.org/10.1051/jphys:019870048070108500} {\bibfield
   {journal} {\bibinfo  {journal} {J. Phys. France}\ }\textbf {\bibinfo
  {volume} {48}},\ \bibinfo {pages} {1085} (\bibinfo {year}
  {1987})}\BibitemShut {NoStop}%
\bibitem [{\citenamefont {Lubensky}\ and\ \citenamefont
  {Prost}(1992)}]{lubensky1992orientational}%
  \BibitemOpen
  \bibfield  {author} {\bibinfo {author} {\bibfnamefont {T.~C.}\ \bibnamefont
  {Lubensky}}\ and\ \bibinfo {author} {\bibfnamefont {J.}~\bibnamefont
  {Prost}},\ }\bibfield  {title} {\bibinfo {title} {Orientational order and
  vesicle shape},\ }\href {https://doi.org/10.1051/jp2:1992133} {\bibfield
  {journal} {\bibinfo  {journal} {Journal de Physique II}\ }\textbf {\bibinfo
  {volume} {2}},\ \bibinfo {pages} {371} (\bibinfo {year} {1992})}\BibitemShut
  {NoStop}%
\bibitem [{\citenamefont {Park}\ and\ \citenamefont
  {Lubensky}(1996)}]{park1996topological}%
  \BibitemOpen
  \bibfield  {author} {\bibinfo {author} {\bibfnamefont {J.-M.}\ \bibnamefont
  {Park}}\ and\ \bibinfo {author} {\bibfnamefont {T.~C.}\ \bibnamefont
  {Lubensky}},\ }\bibfield  {title} {\bibinfo {title} {Topological defects on
  fluctuating surfaces: General properties and the kosterlitz-thouless
  transition},\ }\href {https://doi.org/10.1103/PhysRevE.53.2648} {\bibfield
  {journal} {\bibinfo  {journal} {Phys. Rev. E}\ }\textbf {\bibinfo {volume}
  {53}},\ \bibinfo {pages} {2648} (\bibinfo {year} {1996})}\BibitemShut
  {NoStop}%
\bibitem [{\citenamefont {Bowick}\ and\ \citenamefont
  {Giomi}(2009)}]{bowick2009two}%
  \BibitemOpen
  \bibfield  {author} {\bibinfo {author} {\bibfnamefont {M.~J.}\ \bibnamefont
  {Bowick}}\ and\ \bibinfo {author} {\bibfnamefont {L.}~\bibnamefont {Giomi}},\
  }\bibfield  {title} {\bibinfo {title} {Two-dimensional matter: order,
  curvature and defects},\ }\href {https://doi.org/10.1080/00018730903043166}
  {\bibfield  {journal} {\bibinfo  {journal} {Advances in Physics}\ }\textbf
  {\bibinfo {volume} {58}},\ \bibinfo {pages} {449} (\bibinfo {year} {2009})},\
  \Eprint {https://arxiv.org/abs/https://doi.org/10.1080/00018730903043166}
  {https://doi.org/10.1080/00018730903043166} \BibitemShut {NoStop}%
\bibitem [{\citenamefont {Vafa}\ and\ \citenamefont
  {Mahadevan}(2021)}]{vafa2021active}%
  \BibitemOpen
  \bibfield  {author} {\bibinfo {author} {\bibfnamefont {F.}~\bibnamefont
  {Vafa}}\ and\ \bibinfo {author} {\bibfnamefont {L.}~\bibnamefont
  {Mahadevan}},\ }\href@noop {} {\bibinfo {title} {Active nematic defects and
  epithelial morphogenesis}} (\bibinfo {year} {2021}),\ \Eprint
  {https://arxiv.org/abs/2105.01067} {arXiv:2105.01067 [cond-mat.soft]}
  \BibitemShut {NoStop}%
\bibitem [{\citenamefont {Vitelli}\ and\ \citenamefont
  {Turner}(2004)}]{vitelli2004anomalous}%
  \BibitemOpen
  \bibfield  {author} {\bibinfo {author} {\bibfnamefont {V.}~\bibnamefont
  {Vitelli}}\ and\ \bibinfo {author} {\bibfnamefont {A.~M.}\ \bibnamefont
  {Turner}},\ }\bibfield  {title} {\bibinfo {title} {Anomalous coupling between
  topological defects and curvature},\ }\href
  {https://doi.org/10.1103/PhysRevLett.93.215301} {\bibfield  {journal}
  {\bibinfo  {journal} {Phys. Rev. Lett.}\ }\textbf {\bibinfo {volume} {93}},\
  \bibinfo {pages} {215} (\bibinfo {year} {2004})}\BibitemShut {NoStop}%
\bibitem [{\citenamefont {Wen}\ and\ \citenamefont {Zee}(1992)}]{wen1992shift}%
  \BibitemOpen
  \bibfield  {author} {\bibinfo {author} {\bibfnamefont {X.-G.}\ \bibnamefont
  {Wen}}\ and\ \bibinfo {author} {\bibfnamefont {A.}~\bibnamefont {Zee}},\
  }\bibfield  {title} {\bibinfo {title} {Shift and spin vector: New topological
  quantum numbers for the hall fluids},\ }\href@noop {} {\bibfield  {journal}
  {\bibinfo  {journal} {Physical review letters}\ }\textbf {\bibinfo {volume}
  {69}},\ \bibinfo {pages} {953} (\bibinfo {year} {1992})}\BibitemShut
  {NoStop}%
\bibitem [{\citenamefont {Biswas}\ and\ \citenamefont
  {Son}(2016)}]{biswas2016fractional}%
  \BibitemOpen
  \bibfield  {author} {\bibinfo {author} {\bibfnamefont {R.~R.}\ \bibnamefont
  {Biswas}}\ and\ \bibinfo {author} {\bibfnamefont {D.~T.}\ \bibnamefont
  {Son}},\ }\bibfield  {title} {\bibinfo {title} {Fractional charge and
  inter-landau--level states at points of singular curvature},\ }\href@noop {}
  {\bibfield  {journal} {\bibinfo  {journal} {Proceedings of the National
  Academy of Sciences}\ }\textbf {\bibinfo {volume} {113}},\ \bibinfo {pages}
  {8636} (\bibinfo {year} {2016})}\BibitemShut {NoStop}%
\bibitem [{\citenamefont {Zhang}\ and\ \citenamefont
  {Nelson}(2022)}]{zhang2022fractional}%
  \BibitemOpen
  \bibfield  {author} {\bibinfo {author} {\bibfnamefont {G.~H.}\ \bibnamefont
  {Zhang}}\ and\ \bibinfo {author} {\bibfnamefont {D.~R.}\ \bibnamefont
  {Nelson}},\ }\href@noop {} {\bibinfo {title} {Fractional defect charges in
  $p$-atic liquid crystals on cones}} (\bibinfo {year} {2022}),\ \Eprint
  {https://arxiv.org/abs/2201.11201} {arXiv:2201.11201 [cond-mat.soft]}
  \BibitemShut {NoStop}%
\bibitem [{\citenamefont {Gauss}(1959)}]{gauss1822on}%
  \BibitemOpen
  \bibfield  {author} {\bibinfo {author} {\bibfnamefont {C.~F.}\ \bibnamefont
  {Gauss}},\ }\bibfield  {title} {\bibinfo {title} {On conformal
  representations},\ }in\ \href@noop {} {\emph {\bibinfo {booktitle} {A Source
  Book in Mathematics}}},\ \bibinfo {editor} {edited by\ \bibinfo {editor}
  {\bibfnamefont {D.~E.}\ \bibnamefont {Smith}}}\ (\bibinfo  {publisher}
  {Dover},\ \bibinfo {year} {1959})\ p.\ \bibinfo {pages}
  {463–475}\BibitemShut {NoStop}%
\bibitem [{\citenamefont {David}(2004)}]{david2004geometry}%
  \BibitemOpen
  \bibfield  {author} {\bibinfo {author} {\bibfnamefont {F.}~\bibnamefont
  {David}},\ }\bibfield  {title} {\bibinfo {title} {Geometry and field theory
  of random surfaces and membranes},\ }in\ \href {https://doi.org/10.1142/5473}
  {\emph {\bibinfo {booktitle} {Statistical Mechanics of Membranes and
  Surfaces}}},\ \bibinfo {editor} {edited by\ \bibinfo {editor} {\bibfnamefont
  {D.}~\bibnamefont {Nelson}}, \bibinfo {editor} {\bibfnamefont
  {T.}~\bibnamefont {Piran}},\ and\ \bibinfo {editor} {\bibfnamefont
  {S.}~\bibnamefont {Weinberg}}}\ (\bibinfo  {publisher} {World Scientific},\
  \bibinfo {year} {2004})\ \bibinfo {edition} {2nd}\ ed.,\ Chap.~\bibinfo
  {chapter} {7}, pp.\ \bibinfo {pages} {149--209}\BibitemShut {NoStop}%
\bibitem [{\citenamefont {Turner}\ \emph {et~al.}(2010)\citenamefont {Turner},
  \citenamefont {Vitelli},\ and\ \citenamefont {Nelson}}]{turner2010vortices}%
  \BibitemOpen
  \bibfield  {author} {\bibinfo {author} {\bibfnamefont {A.~M.}\ \bibnamefont
  {Turner}}, \bibinfo {author} {\bibfnamefont {V.}~\bibnamefont {Vitelli}},\
  and\ \bibinfo {author} {\bibfnamefont {D.~R.}\ \bibnamefont {Nelson}},\
  }\bibfield  {title} {\bibinfo {title} {Vortices on curved surfaces},\ }\href
  {https://doi.org/10.1103/RevModPhys.82.1301} {\bibfield  {journal} {\bibinfo
  {journal} {Rev. Mod. Phys.}\ }\textbf {\bibinfo {volume} {82}},\ \bibinfo
  {pages} {1301} (\bibinfo {year} {2010})}\BibitemShut {NoStop}%
\bibitem [{\citenamefont {Bowick}\ \emph {et~al.}(2004)\citenamefont {Bowick},
  \citenamefont {Nelson},\ and\ \citenamefont
  {Travesset}}]{bowick2004curvature}%
  \BibitemOpen
  \bibfield  {author} {\bibinfo {author} {\bibfnamefont {M.}~\bibnamefont
  {Bowick}}, \bibinfo {author} {\bibfnamefont {D.~R.}\ \bibnamefont {Nelson}},\
  and\ \bibinfo {author} {\bibfnamefont {A.}~\bibnamefont {Travesset}},\
  }\bibfield  {title} {\bibinfo {title} {Curvature-induced defect unbinding in
  toroidal geometries},\ }\href {https://doi.org/10.1103/PhysRevE.69.041102}
  {\bibfield  {journal} {\bibinfo  {journal} {Phys. Rev. E}\ }\textbf {\bibinfo
  {volume} {69}},\ \bibinfo {pages} {041102} (\bibinfo {year}
  {2004})}\BibitemShut {NoStop}%
\bibitem [{\citenamefont {Nakahara}(2003)}]{nakahara2003geometry}%
  \BibitemOpen
  \bibfield  {author} {\bibinfo {author} {\bibfnamefont {M.}~\bibnamefont
  {Nakahara}},\ }\href@noop {} {\emph {\bibinfo {title} {Geometry, Topology and
  Physics}}}\ (\bibinfo  {publisher} {Taylor \& Francis},\ \bibinfo {year}
  {2003})\BibitemShut {NoStop}%
\bibitem [{\citenamefont {Giomi}\ \emph
  {et~al.}(2021{\natexlab{a}})\citenamefont {Giomi}, \citenamefont {Toner},\
  and\ \citenamefont {Sarkar}}]{giomi2021hydrodynamic}%
  \BibitemOpen
  \bibfield  {author} {\bibinfo {author} {\bibfnamefont {L.}~\bibnamefont
  {Giomi}}, \bibinfo {author} {\bibfnamefont {J.}~\bibnamefont {Toner}},\ and\
  \bibinfo {author} {\bibfnamefont {N.}~\bibnamefont {Sarkar}},\ }\href@noop {}
  {\bibinfo {title} {Hydrodynamic theory of $p-$atic liquid crystals}}
  (\bibinfo {year} {2021}{\natexlab{a}}),\ \Eprint
  {https://arxiv.org/abs/2106.11957} {arXiv:2106.11957 [cond-mat.soft]}
  \BibitemShut {NoStop}%
\bibitem [{\citenamefont {Giomi}\ \emph
  {et~al.}(2021{\natexlab{b}})\citenamefont {Giomi}, \citenamefont {Toner},\
  and\ \citenamefont {Sarkar}}]{giomi2021longranged}%
  \BibitemOpen
  \bibfield  {author} {\bibinfo {author} {\bibfnamefont {L.}~\bibnamefont
  {Giomi}}, \bibinfo {author} {\bibfnamefont {J.}~\bibnamefont {Toner}},\ and\
  \bibinfo {author} {\bibfnamefont {N.}~\bibnamefont {Sarkar}},\ }\href@noop {}
  {\bibinfo {title} {Long-ranged order and flow alignment in sheared $p-$atic
  liquid crystals}} (\bibinfo {year} {2021}{\natexlab{b}}),\ \Eprint
  {https://arxiv.org/abs/2111.04720} {arXiv:2111.04720 [cond-mat.soft]}
  \BibitemShut {NoStop}%
\bibitem [{\citenamefont {Vafa}\ \emph {et~al.}(2020)\citenamefont {Vafa},
  \citenamefont {Bowick}, \citenamefont {Marchetti},\ and\ \citenamefont
  {Shraiman}}]{vafa2020multi-defect}%
  \BibitemOpen
  \bibfield  {author} {\bibinfo {author} {\bibfnamefont {F.}~\bibnamefont
  {Vafa}}, \bibinfo {author} {\bibfnamefont {M.~J.}\ \bibnamefont {Bowick}},
  \bibinfo {author} {\bibfnamefont {M.~C.}\ \bibnamefont {Marchetti}},\ and\
  \bibinfo {author} {\bibfnamefont {B.~I.}\ \bibnamefont {Shraiman}},\
  }\href@noop {} {\bibinfo {title} {Multi-defect dynamics in active nematics}}
  (\bibinfo {year} {2020}),\ \Eprint {https://arxiv.org/abs/2007.02947}
  {arXiv:2007.02947 [cond-mat.soft]} \BibitemShut {NoStop}%
\bibitem [{\citenamefont {Seung}\ and\ \citenamefont
  {Nelson}(1988)}]{seung1988defects}%
  \BibitemOpen
  \bibfield  {author} {\bibinfo {author} {\bibfnamefont {H.~S.}\ \bibnamefont
  {Seung}}\ and\ \bibinfo {author} {\bibfnamefont {D.~R.}\ \bibnamefont
  {Nelson}},\ }\bibfield  {title} {\bibinfo {title} {Defects in flexible
  membranes with crystalline order},\ }\href
  {https://doi.org/10.1103/PhysRevA.38.1005} {\bibfield  {journal} {\bibinfo
  {journal} {Phys. Rev. A}\ }\textbf {\bibinfo {volume} {38}},\ \bibinfo
  {pages} {1005} (\bibinfo {year} {1988})}\BibitemShut {NoStop}%
\bibitem [{\citenamefont {Selinger}(2015)}]{selinger2015introduction}%
  \BibitemOpen
  \bibfield  {author} {\bibinfo {author} {\bibfnamefont {J.~V.}\ \bibnamefont
  {Selinger}},\ }\href@noop {} {\emph {\bibinfo {title} {Introduction to the
  theory of soft matter: from ideal gases to liquid crystals}}}\ (\bibinfo
  {publisher} {Springer},\ \bibinfo {year} {2015})\BibitemShut {NoStop}%
\bibitem [{\citenamefont {Zwillinger}(2018)}]{zwillinger2018crc}%
  \BibitemOpen
  \bibfield  {author} {\bibinfo {author} {\bibfnamefont {D.}~\bibnamefont
  {Zwillinger}},\ }\href@noop {} {\emph {\bibinfo {title} {CRC standard
  mathematical tables and formulas}}}\ (\bibinfo  {publisher} {chapman and
  hall/CRC},\ \bibinfo {year} {2018})\BibitemShut {NoStop}%
\bibitem [{\citenamefont {Broyden}(1970)}]{broyden1970convergence}%
  \BibitemOpen
  \bibfield  {author} {\bibinfo {author} {\bibfnamefont {C.~G.}\ \bibnamefont
  {Broyden}},\ }\bibfield  {title} {\bibinfo {title} {The convergence of a
  class of double-rank minimization algorithms 1. general considerations},\
  }\href@noop {} {\bibfield  {journal} {\bibinfo  {journal} {IMA Journal of
  Applied Mathematics}\ }\textbf {\bibinfo {volume} {6}},\ \bibinfo {pages}
  {76} (\bibinfo {year} {1970})}\BibitemShut {NoStop}%
\bibitem [{\citenamefont {Fletcher}(1970)}]{fletcher1970new}%
  \BibitemOpen
  \bibfield  {author} {\bibinfo {author} {\bibfnamefont {R.}~\bibnamefont
  {Fletcher}},\ }\bibfield  {title} {\bibinfo {title} {A new approach to
  variable metric algorithms},\ }\href@noop {} {\bibfield  {journal} {\bibinfo
  {journal} {The computer journal}\ }\textbf {\bibinfo {volume} {13}},\
  \bibinfo {pages} {317} (\bibinfo {year} {1970})}\BibitemShut {NoStop}%
\bibitem [{\citenamefont {Goldfarb}(1970)}]{goldfarb1970family}%
  \BibitemOpen
  \bibfield  {author} {\bibinfo {author} {\bibfnamefont {D.}~\bibnamefont
  {Goldfarb}},\ }\bibfield  {title} {\bibinfo {title} {A family of
  variable-metric methods derived by variational means},\ }\href@noop {}
  {\bibfield  {journal} {\bibinfo  {journal} {Mathematics of computation}\
  }\textbf {\bibinfo {volume} {24}},\ \bibinfo {pages} {23} (\bibinfo {year}
  {1970})}\BibitemShut {NoStop}%
\bibitem [{\citenamefont {Shanno}(1970)}]{shanno1970conditioning}%
  \BibitemOpen
  \bibfield  {author} {\bibinfo {author} {\bibfnamefont {D.~F.}\ \bibnamefont
  {Shanno}},\ }\bibfield  {title} {\bibinfo {title} {Conditioning of
  quasi-newton methods for function minimization},\ }\href@noop {} {\bibfield
  {journal} {\bibinfo  {journal} {Mathematics of computation}\ }\textbf
  {\bibinfo {volume} {24}},\ \bibinfo {pages} {647} (\bibinfo {year}
  {1970})}\BibitemShut {NoStop}%
\bibitem [{\citenamefont {Moore}\ and\ \citenamefont
  {P{\'e}rez-Garrido}(1999)}]{moore1999absence}%
  \BibitemOpen
  \bibfield  {author} {\bibinfo {author} {\bibfnamefont {M.}~\bibnamefont
  {Moore}}\ and\ \bibinfo {author} {\bibfnamefont {A.}~\bibnamefont
  {P{\'e}rez-Garrido}},\ }\bibfield  {title} {\bibinfo {title} {Absence of a
  finite-temperature melting transition in the classical two-dimensional
  one-component plasma},\ }\href@noop {} {\bibfield  {journal} {\bibinfo
  {journal} {Physical review letters}\ }\textbf {\bibinfo {volume} {82}},\
  \bibinfo {pages} {4078} (\bibinfo {year} {1999})}\BibitemShut {NoStop}%
\bibitem [{\citenamefont {Bowick}\ \emph {et~al.}(2000)\citenamefont {Bowick},
  \citenamefont {Nelson},\ and\ \citenamefont
  {Travesset}}]{bowick2000interacting}%
  \BibitemOpen
  \bibfield  {author} {\bibinfo {author} {\bibfnamefont {M.~J.}\ \bibnamefont
  {Bowick}}, \bibinfo {author} {\bibfnamefont {D.~R.}\ \bibnamefont {Nelson}},\
  and\ \bibinfo {author} {\bibfnamefont {A.}~\bibnamefont {Travesset}},\
  }\bibfield  {title} {\bibinfo {title} {Interacting topological defects on
  frozen topographies},\ }\href@noop {} {\bibfield  {journal} {\bibinfo
  {journal} {Physical Review B}\ }\textbf {\bibinfo {volume} {62}},\ \bibinfo
  {pages} {8738} (\bibinfo {year} {2000})}\BibitemShut {NoStop}%
\bibitem [{\citenamefont {Bruinsma}\ \emph {et~al.}(1982)\citenamefont
  {Bruinsma}, \citenamefont {Halperin},\ and\ \citenamefont
  {Zippelius}}]{bruinsma1982motion}%
  \BibitemOpen
  \bibfield  {author} {\bibinfo {author} {\bibfnamefont {R.}~\bibnamefont
  {Bruinsma}}, \bibinfo {author} {\bibfnamefont {B.}~\bibnamefont {Halperin}},\
  and\ \bibinfo {author} {\bibfnamefont {A.}~\bibnamefont {Zippelius}},\
  }\bibfield  {title} {\bibinfo {title} {Motion of defects and stress
  relaxation in two-dimensional crystals},\ }\href@noop {} {\bibfield
  {journal} {\bibinfo  {journal} {Physical Review B}\ }\textbf {\bibinfo
  {volume} {25}},\ \bibinfo {pages} {579} (\bibinfo {year} {1982})}\BibitemShut
  {NoStop}%
\end{thebibliography}%
	
\end{document}